\documentclass[sigconf]{acmart} 
\usepackage{algorithm}
\usepackage{algorithmic}

\usepackage{makecell}
\usepackage{multirow}
\usepackage{array}
\usepackage{amsmath}
\usepackage{amsthm}
\usepackage{booktabs}
\usepackage{graphicx}
\usepackage{flushend}
\usepackage{bm}
\usepackage{tabularx}
\usepackage{subcaption}
\usepackage[most]{tcolorbox}
\usepackage{enumitem}
\setlist[itemize]{leftmargin=*, nosep}
\usepackage[table]{xcolor}
\usepackage{xspace}

\newcommand{\model}{SemaCDR\xspace}

\newcommand{\eg}{\emph{e.g.,}\xspace}
\newcommand{\ie}{\emph{i.e.,}\xspace}

\AtBeginDocument{%
  }

\copyrightyear{2026}
\acmYear{2026}
\setcopyright{cc}
\setcctype{by}
\acmConference[WWW '26]{Proceedings of the ACM Web Conference 2026}{April 13--17, 2026}{Dubai, United Arab Emirates}
\acmBooktitle{Proceedings of the ACM Web Conference 2026 (WWW '26), April 13--17, 2026, Dubai, United Arab Emirates}
\acmISBN{979-8-4007-2307-0/2026/04}
\acmDOI{10.1145/3774904.3792651} 

\renewcommand{\shortauthors}{Zhang et al.} 
\begin{document}

\title{SemaCDR: LLM-Powered Transferable Semantics for Cross-Domain Sequential Recommendation}

\author{Chunxu Zhang}
\authornote{Also with the Key Laboratory of Symbolic Computation and Knowledge Engineering of Ministry of Education, Jilin University.}
\affiliation{%
  \institution{Jilin University}
  \city{Changchun}
  \country{China} \\
  \institution{Hong Kong Polytechnic University}
  \city{Hong Kong}
  \country{China}
}
\email{zhangchunxu@jlu.edu.cn} 

\author{Shanqiang Huang}
\affiliation{%
  \institution{Jilin University}
  \city{Changchun}
  \country{China}
}
\email{huangsq2022@mails.jlu.edu.cn} 

\author{Zijian Zhang} 
\authornotemark[1]
\authornote{Corresponding authors.}
\affiliation{
  \institution{Jilin University}
  \city{Changchun}
  \country{China} \\
  \institution{The Chinese University of Hong Kong}
   \city{Hong Kong}
   \country{China}
}
\email{zhangzijian@jlu.edu.cn}

\author{Jiahong Liu}
\affiliation{%
 \institution{The Chinese University of Hong Kong}
   \city{Hong Kong}
   \country{China}
}
\email{jiahong.liu21@gmail.com	}

\author{Linsong Yu}
\affiliation{%
 \institution{Jilin University}
   \city{Changchun}
   \country{China}
}
\email{lsyu9923@mails.jlu.edu.cn}

\author{Ruiqi Wan}
\affiliation{%
 \institution{Jilin University}
   \city{Changchun}
   \country{China}
}
\email{wanrq2224@mails.jlu.edu.cn}

\author{Bo Yang}
\authornotemark[1]
\authornotemark[2]
\affiliation{%
 \institution{Jilin University}
   \city{Changchun}
   \country{China}
}
\email{ybo@jlu.edu.cn}

\author{Irwin King}
\affiliation{%
 \institution{The Chinese University of Hong Kong}
   \city{Hong Kong}
   \country{China}
}
\email{king@cse.cuhk.edu.hk}

\renewcommand{\shortauthors}{Chunxu Zhang et al.}

\begin{abstract}
  Cross-domain recommendation (CDR) addresses the data sparsity and cold-start problems in the target domain by leveraging knowledge from data-rich source domains. However, existing CDR methods often rely on domain-specific features or identifiers that lack transferability across different domains, limiting their ability to capture inter-domain semantic patterns. To overcome this, we propose \model, a semantics-driven framework for cross-domain sequential recommendation that leverages large language models (LLMs) to construct a unified semantic space. \model creates multi-view item features by integrating LLM-generated domain-agnostic semantics with domain-specific content, aligned by contrastive regularization. \model systematically creates LLM-generated domain-specific and domain-agnostic semantics, and employs adaptive fusion to generate unified preference representations. Furthermore, it aligns cross-domain behavior sequences with an adaptive fusion mechanism to synthesize interaction sequences from source, target, and mixed domains. Extensive experiments on real-world datasets show that \model consistently outperforms state-of-the-art baselines, demonstrating its effectiveness in capturing coherent intra-domain patterns while facilitating knowledge transfer across domains. Our code is available online\footnote{\url{https://github.com/huangshanqiang/SemaCDR}}.

  
  
\end{abstract}



\maketitle

\section{Introduction}
In the field of recommender systems, Cross-Domain Recommendation (CDR) is a critical solution for systems that lack sufficient interaction history. 
The fundamental premise of CDR is the assumption that a user's latent preferences are largely consistent across different domains. 
By leveraging rich interaction data and learned interest representations from a data-rich source domain, CDR effectively enhances the modeling of interaction preferences in the target domain. 
This knowledge transfer provides vital predictive signals for cold-start users and items with minimal history~\cite{park2024pacer,zhang2025comprehensive,zhu2021cross,man2017cross,zang2022survey,cao2023towards}, enhancing the robustness and completeness of user interest modeling and improving recommendation accuracy and coverage.



Existing research in CDR has generally evolved into three dominant methodological streams. 
The first involves mapping-based approaches~\cite{man2017cross,zhu2021transfer}, which learn explicit transformation functions to align user or item representations between the source and target domains, thereby enabling direct feature transfer. 
The second focuses on fusion-based approaches~\cite{guo2021gcn,li2023preference,liu2025spottrip}, where heterogeneous signals from multiple domains are jointly modeled at various levels of granularity, \eg via shared factorization or graph structures, to capture richer cross-domain interactions. 
The third, and most recent stream, explores Large Language Models (LLM)-based approaches~\cite{huang2025large,chen2024large,wu2024survey,liu2024large,lyu2023llm,wang2024towards}, where large language models are either employed as expressive feature encoders to enrich item representations or directly adapted as transferable recommendation models to facilitate knowledge transfer across domains. 


\begin{figure}[!t]
\setlength{\abovecaptionskip}{1mm}
\setlength{\belowcaptionskip}{-1mm}
{{\includegraphics[width=1\linewidth]{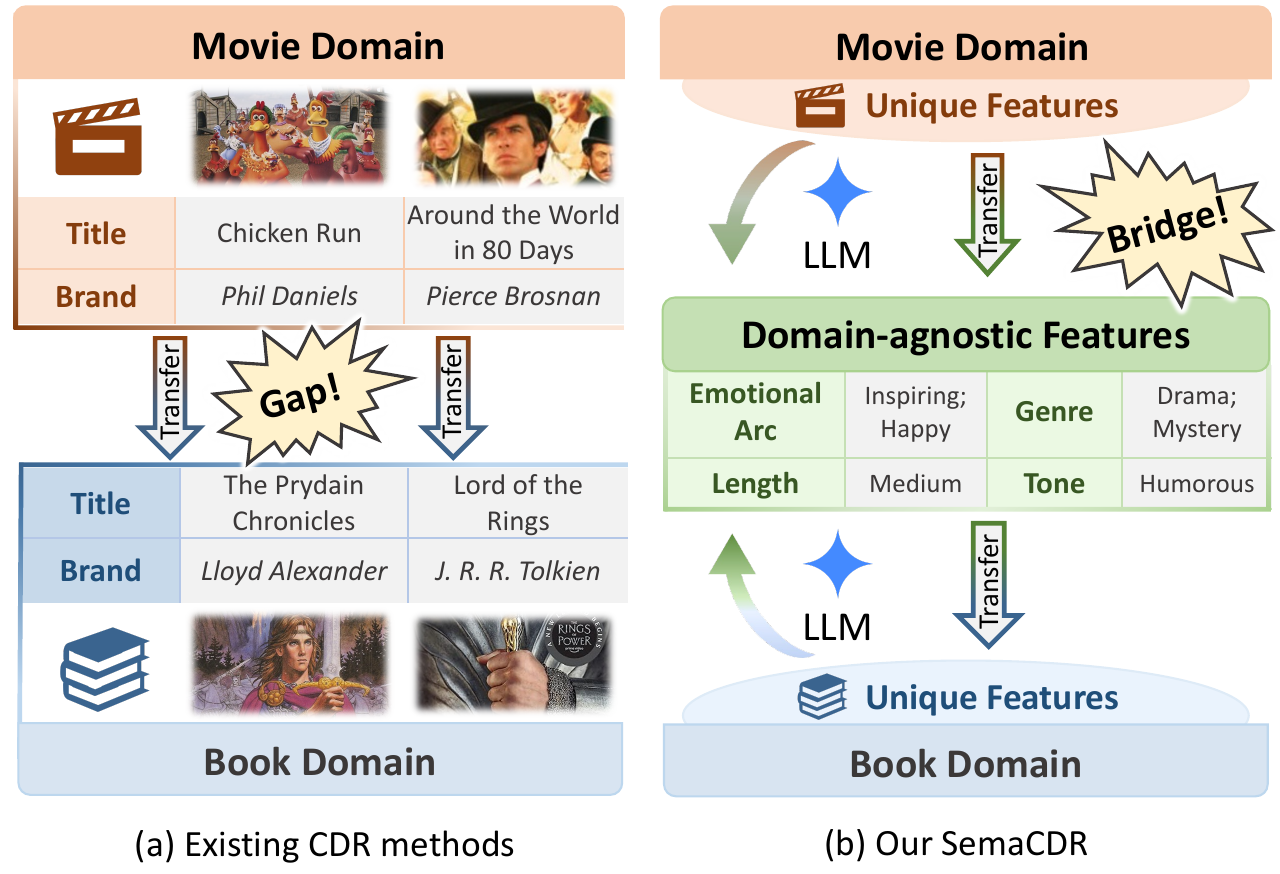}}}
\caption{Comparison of existing CDR methods and our proposed \model for cross-domain knowledge transfer.}
    \label{fig:motivation}
\end{figure}

Despite the continuous technological advancements across existing methods, a fundamental semantic bottleneck persists: \textit{the inability to capture transferable semantic patterns that consistently characterize user preferences across domains}. 
As shown in Figure~\ref{fig:motivation} (a), existing CDR models primarily rely on domain-specific features, such as movie and book titles or brands. 
While these features are crucial for intra-domain prediction, they often struggle to facilitate effective knowledge transfer. 
We identify that the core challenge of cross-domain recommendation is to move beyond superficial transfer and achieve effective semantic-level migration.
This requires a robust intermediary capable of converting low-level, domain-bound feature details into high-level, domain-agnostic knowledge. 
LLMs possess rich semantic understanding and broad world knowledge, naturally providing a promising means to generate unified, transferable representations that capture cross-domain user preferences.

\begin{figure*}[!t]
\setlength{\abovecaptionskip}{1mm}
{{\includegraphics[width=0.9\linewidth]{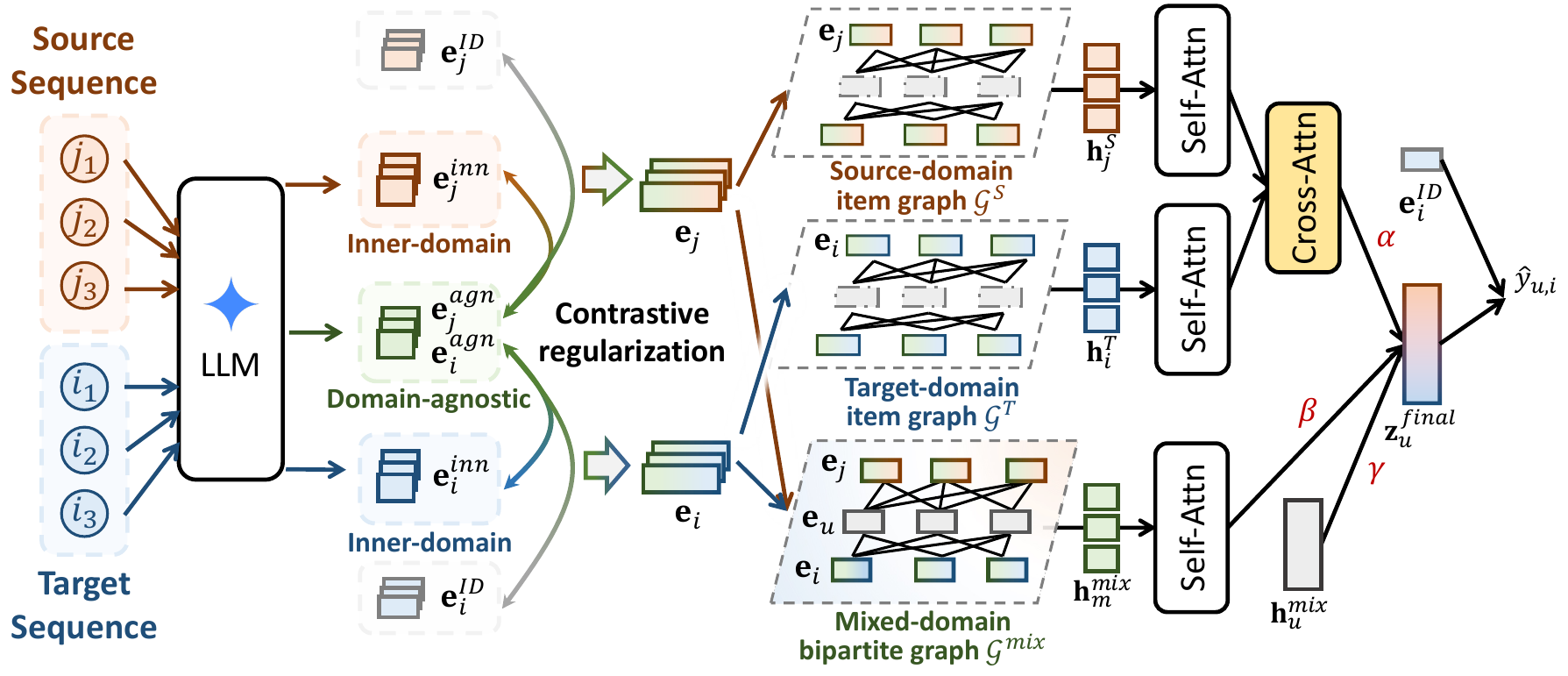}}}
\caption{Overview of \model\ for cross-domain sequential recommendation. The multi-view semantic learning module constructs and aligns domain-specific and domain-agnostic item embeddings, which feed into the cross-domain behavior fusion module to encode sequential user interactions. The adaptive fusion prediction module then integrates these embeddings with mixed-domain and user representations to produce unified preference predictions, bridging cross-domain knowledge.}
    \label{fig:framework}
\end{figure*}
In this work, we introduce \model, a cross-domain sequential recommendation framework that leverages LLMs to construct a unified semantic space capturing latent patterns in user preferences. Rather than focusing solely on source-to-target feature alignment, \model aims to systematically integrate multi-dimensional semantics that consistently shape user behavior across domains. (i) domain-specific semantics, which integrate identifiers and intrinsic content features to capture item uniqueness and enrich their characterization, and (ii) domain-agnostic latent semantics, which uncover deeper cross-domain commonalities for effective knowledge transfer as illustrated in Figure~\ref{fig:motivation} (b). Building on these representations, a cross-domain behavior fusion module integrates user interaction sequences from the source and target domains into cohesive sequential embeddings that encode inter-domain dynamics. Finally, an adaptive fusion prediction mechanism combines these sequential embeddings with user and mixed-domain representations to generate unified preference estimations. This holistic design allows \model\ to exploit complementary semantics, align behavioral signals, and flexibly adapt to diverse cross-domain recommendation scenarios. The main contributions can be summarized as follows,
\begin{itemize}
    \item We propose \model, a semantics-driven framework for cross-domain sequential recommendation that leverages LLMs to construct a unified semantic space, enabling the capture of latent patterns that consistently govern user preferences across heterogeneous domains.
    \item We introduce a multi-view semantic learning mechanism to capture complementary item representations and, together with the cross-domain behavior and adaptive fusion strategies, integrate sequential user interactions, enhancing the model’s ability to capture coherent intra-domain patterns while effectively transferring knowledge across domains.
    \item Extensive experiments on real-world datasets demonstrate that \model consistently outperforms state-of-the-art baselines across diverse metrics, with analyses confirming the effectiveness of its design and its compatibility with existing cross-domain recommendation architectures.
\end{itemize}






\section{Methodology}
\subsection{Problem Formulation}
We consider a Cross-Domain Sequential Recommendation (CDSR) scenario consisting of a \textit{source domain} $S$ and a \textit{target domain} $T$, which share a common user set $\mathcal{U}$. Each domain is associated with a distinct item set, denoted as $\mathcal{I}^S$ for the source and $\mathcal{I}^T$ for the target. For each user $u \in \mathcal{U}$, we define a chronological interaction sequence $\mathcal{S}^S_u = [j_{u,1}, j_{u,2}, \ldots, j_{u,n_S}]$ in the source domain and $\mathcal{S}^T_u = [i_{u,1}, i_{u,2}, \ldots, i_{u,n_T}]$ in the target domain, where each element denotes an item the user has interacted with in temporal order. These sequences capture users’ dynamic behavioral patterns within and across domains, providing the foundation for preference transfer and temporal modeling. In addition, each item may be accompanied by raw descriptive data, such as textual metadata and reviews. We denote these raw features as $\mathbf{d}^S_j$ for source items $j \in \mathcal{I}^S$ and $\mathbf{d}^T_i$ for target items $i \in \mathcal{I}^T$. These heterogeneous signals serve as auxiliary information to enhance item characterization. The objective of CDSR is to exploit both sequential interaction data and auxiliary raw features from the source and target domains to improve user preference modeling in the target domain. Formally, the task is to learn a predictive function
\begin{equation}
    f: \mathcal{U} \times \mathcal{I}^T \rightarrow \mathbb{R},
\end{equation}
which estimates the preference score $\hat{y}_{u, i} = f(u, i)$ for user $u \in \mathcal{U}$ and target item $i \in \mathcal{I}^T$, by effectively leveraging both domain-specific interaction sequences and the associated raw features.

\subsection{Framework Overview}
As illustrated in {Figure}~\ref{fig:framework}, \model\ consists of three components: multi-view semantic learning, cross-domain behavior fusion, and adaptive fusion prediction. 
The semantic module aligns domain-specific and domain-agnostic item representations via contrastive learning and graph modeling to establish a transferable foundation. Subsequently, the behavior fusion module integrates cross-domain interaction sequences to produce enhanced embeddings. These are finally combined with user and mixed-domain contexts through an adaptive fusion strategy for preference inference, enabling \model\ to effectively exploit complementary semantics and align cross-domain behaviors.

\subsection{Multi-view Semantic Learning}
\subsubsection{Domain-specific semantic embedding}
To capture the unique semantics of items in each domain, we construct domain-specific semantic embeddings. These embeddings serve as the foundation for modeling item individuality in the target domain, helping user preferences to be better aligned with domain-specific item characteristics. Concretely, for the target domain $T$, each item $i \in \mathcal{I}^T$ is represented in two complementary ways: 

\textbf{(1) Identifier-based embedding.} Each item $i \in \mathcal{I}^T$ is assigned a unique identifier, mapped via a trainable embedding table to a latent vector $\mathbf{e}^{ID}_i \in \mathbb{R}^{d_1}$, where $d_1$ is the embedding dimension. This embedding ensures item distinctiveness within the domain.

\textbf{(2) Inner-domain semantic representation.} Beyond identifiers, each item’s raw features $\mathbf{d}^T_i$ are processed by prompting an LLM to produce enriched semantic descriptions, which are then encoded into continuous embeddings $\mathbf{e}^{inn}_i \in \mathbb{R}^{d_2}$ using a pre-trained language model. This embedding augments identifier representations with domain-specific semantics from item content. The same procedure applies to source-domain items $\mathcal{I}^S$.


\subsubsection{Domain-agnostic semantic embedding}
Domain-agnostic semantics aim to extract transferable representations that summarize shared characteristics across domains, bridging knowledge transfer and reducing source–target discrepancies. For example, in the Movie–Book scenario, we prompt LLMs with tailored instructions to generate domain-agnostic semantic categories, typically including Genre and Target Audience, with finer-grained subcategories such as Drama and Historical. Detailed examples and prompt templates are provided in Appendix~\ref{app:llm}.


Formally, we consider a set of $K$ high-level semantic categories $\mathcal{C} = \{c_1, c_2, \dots, c_K\}$, derived from both source and target items through an LLM with carefully designed prompts. Each high-level category $c_k$ is further divided into a set of fine-grained subcategories $\mathcal{V}_{c_k} = \{v_{1}, \dots, v_{m_k}\}$, where $m_k$ may vary across categories. For an item $m \in \mathcal{I}^S \cup \mathcal{I}^T$, its domain-agnostic embedding is constructed by selecting one or more subcategories from each high-level category and combining the corresponding subcategory embeddings,
\begin{equation}
   \textstyle \mathbf{e}_m^{{agn}} = \bigoplus_{k=1}^{K} \bigoplus_{v \in \mathcal{V}_m(c_k)} \mathbf{e}(v),
\end{equation}
where $\mathcal{V}_m(c_k) \subseteq \mathcal{V}_{c_k}$ denotes the set of subcategories assigned to item $m$ under category $c_k$, $\mathbf{e}(v)$ is the learnable embedding of subcategory $v$, and $\bigoplus$ denotes a generic embedding integration operation such as concatenation or pooling. This construction produces domain-agnostic embeddings that capture cross-domain commonalities, thereby mitigating distributional gaps and supporting effective knowledge transfer.

\subsubsection{Contrastive regularization between item views}
Modeling item relationships is vital for capturing how user preferences generalize across semantically related items. 
Each item has domain-specific embeddings that capture local characteristics and domain-agnostic embeddings that encode transferable semantics. The latter reflect meaningful cross-domain relations, guiding the structuring of domain-specific embeddings to preserve these patterns. To achieve this, we employ a contrastive learning objective that anchors domain-specific embeddings to their domain-agnostic counterparts, enforcing consistent inter-item relationships. Formally, the generic contrastive loss $\mathcal{L}_{\mathrm{c}}(\mathbf{e}^{query}, \mathbf{e}^{key})$ is defined as,
\begin{align}
\mathcal{L}_{\mathrm{c}} &=-
\sum_{i \in \mathcal{I}^T} \log \frac{\exp\big( \mathrm{sim}(\mathbf{e}_i^{query}, \mathbf{e}_i^{key}) / \tau \big)}
{\sum_{i' \in \mathcal{I}^T} \exp\big( \mathrm{sim}(\mathbf{e}_i^{query}, \mathbf{e}_{i'}^{key}) / \tau \big)} \notag \\
&-\sum_{j \in \mathcal{I}^S} \log \frac{\exp\big( \mathrm{sim}(\mathbf{e}_j^{query}, \mathbf{e}_j^{key}) / \tau \big)}
{\sum_{j' \in \mathcal{I}^S} \exp\big( \mathrm{sim}(\mathbf{e}_j^{query}, \mathbf{e}_{j'}^{key}) / \tau \big)},
\end{align}
where $\mathrm{sim}(\cdot,\cdot)$ denotes a similarity function, such as cosine similarity, and $\tau$ is a temperature hyperparameter. 
To effectively align the semantic structures, the total regularization term aggregates this loss across both source and target domains for both view pairs: $\mathcal{L}_{\mathrm{c}}^1(\mathbf{e}^{ID}, \mathbf{e}^{agn})$ and $\mathcal{L}_{\mathrm{c}}^2(\mathbf{e}^{inn}, \mathbf{e}^{agn})$.
This loss aligns the domain-specific embeddings with the relational structure of the domain-agnostic embeddings, promoting consistent semantic organization that enhances downstream preference modeling.

\subsubsection{Graph-based collaborative learning}
To leverage collaborative signals in user–item interactions, which capture behavior patterns and item relationships, \model constructs three graphs: 

\textbf{(1) source-domain item graph $\mathcal{G}^S$}capturing associations among items co-interacted by the same users; 

\textbf{(2) target-domain item graph $\mathcal{G}^T$} constructed similarly; and 

\textbf{(3) mixed-domain bipartite graph $\mathcal{G}^{\mathrm{mix}}$} linking users with items across domains. 

All interactions, including those in $\mathcal{G}^{\mathrm{mix}}$ from both domains, are modeled as equally weighted binary edges. These graphs enable structured propagation of collaborative knowledge.


To exploit the structural information in these graphs, \model employs graph neural networks (GNNs) to propagate collaborative signals among items and users, capturing intra-domain affinities and cross-domain interactions. Node representations are initialized with embeddings from the previous stage: for each item $m \in \mathcal{I}^S \cup \mathcal{I}^T$, $\mathbf{e}_m = \mathbf{e}_m^{ID} \oplus \mathbf{e}_m^{inn} \oplus \mathbf{e}_m^{agn}$, and each user $u \in \mathcal{U}$ by a learnable ID embedding $\mathbf{e}_u$. The GNN updates these nodes to produce enriched node representations reflecting both local and global patterns.
\begin{align}
\mathbf{h}_j^S &= \mathrm{GNN}^S\big(\mathcal{G}^S, \{\mathbf{e}_j \mid j \in \mathcal{I}^S\}\big), \\
\mathbf{h}_i^T &= \mathrm{GNN}^T\big(\mathcal{G}^T, \{\mathbf{e}_i \mid i \in \mathcal{I}^T\}\big), \\
\mathbf{h}_u^{\mathrm{mix}}, \mathbf{h}_m^{\mathrm{mix}}
    &= \mathrm{GNN}^M\Big(\mathcal{G}^{\mathrm{mix}}, 
        \{\mathbf{e}_u, \mathbf{e}_m \mid 
        u \in \mathcal{U}, m \in \mathcal{I}^S \cup \mathcal{I}^T\}\Big).
\end{align}
Through this graph-based collaborative learning, \model effectively leverages user-item interaction patterns to enhance item and user representations, capturing intra-domain structure and bridging cross-domain semantics for improved recommendation modeling.

\subsection{Cross-domain Behavior Fusion}
\subsubsection{Multi-view sequence representation learning}

Sequential patterns in user interactions reveal evolving preferences and offer rich signals for capturing short- and long-term behaviors. In the cross-domain setting, we consider three sequences per user $u \in \mathcal{U}$: \textbf{(1) Source-domain sequence} $\mathcal{S}^S_u$, the user’s interactions in the source domain; \textbf{(2) Target-domain sequence} $\mathcal{S}^T_u$, interactions in the target domain; and \textbf{(3) Mixed-domain sequence} $\mathcal{S}^{\mathrm{mix}}_u$, aggregating interactions from both domains chronologically to capture cross-domain temporal dependencies.

To capture sequential dependencies and contextual information, each sequence is encoded with a Transformer using the graph-enhanced item embeddings from the previous collaborative learning step. For each user $u \in \mathcal{U}$, the sequence representation learning can be formally expressed as,
\begin{align}
\mathbf{z}^S_u &= \mathrm{Transformer}^S(\mathcal{S}^S_u, \{\mathbf{h}_j^S|j \in \mathcal{S}^S_u\}), \\
\mathbf{z}^T_u &= \mathrm{Transformer}^T(\mathcal{S}^T_u, \{\mathbf{h}_i^T|i \in \mathcal{S}^T_u\}), \\
\mathbf{z}^{\mathrm{mix}}_u &= \mathrm{Transformer}^M(\mathcal{S}^{\mathrm{mix}}_u, \{\mathbf{h}_m^{\mathrm{mix}}|m \in \mathcal{S}^{\mathrm{mix}}_u\}).
\end{align}
These multi-view sequence embeddings enable \model to model both domain-specific user behaviors and cross-domain interaction patterns, effectively consolidating sequential information that reflects how preferences evolve and transfer across different domains.

\subsubsection{Cross-attention sequence fusion}

The source- and target-domain sequences individually encode complementary domain-specific dynamics. Fusing these sequences enables the model to leverage cross-domain signals to enhance target-domain representations and improve inter-domain behavioral alignment. To this end, we employ a cross-attention mechanism to merge the source- and target-domain sequence embeddings, enabling complementary dynamics to enhance the refined target-domain sequence. Formally, this fusion can be expressed as,
\begin{equation}
\mathbf{z}_u^{\mathrm{fuse}} = \mathrm{CrossAttn}\big(\mathbf{Q} = \mathbf{z}_u^T, \mathbf{K} = \mathbf{z}_u^S, \mathbf{V} = \mathbf{z}_u^S \big),
\end{equation}
where \(\mathrm{CrossAttn}(\cdot)\) computes attention of each element in the target sequence over the source sequence, producing a fused embedding \(\mathbf{z}_u^{\mathrm{fuse}}\) that integrates complementary cross-domain signals and strengthens inter-domain behavioral modeling.

\subsection{Adaptive Fusion Prediction}

User preferences in cross-domain recommendation are shaped by domain-specific behaviors, cross-domain interactions, and inherent user traits. While the fused target-domain sequence $\mathbf{z}_u^{\mathrm{fuse}}$ captures target-specific patterns, the mixed-domain sequence $\mathbf{z}_u^{\mathrm{mix}}$ encodes overall cross-domain context, and the user embedding $\mathbf{h}_u^{\mathrm{mix}}$ reflects personalized characteristics. To effectively balance these complementary signals, \model\ employs an adaptive weighting fusion strategy to produce a unified sequence representation.
\begin{equation}
    \mathbf{z}_u^{\mathrm{final}} = \alpha \mathbf{z}_u^{\mathrm{fuse}} + \beta \mathbf{z}_u^{\mathrm{mix}} + \gamma \mathbf{h}_u^{\mathrm{mix}},
\end{equation}
where $\alpha, \beta, \gamma$ are learnable parameters weighting each signal. This adaptive fusion produces a representation capturing both cross-domain dynamics and user-specific preferences, which is then used to predict the preference score for target item $i \in \mathcal{I}^T$,
\begin{equation}
\hat{y}_{u,i} = \phi \big( \mathbf{z}^{\mathrm{final}}_u, \mathbf{e}_i^{ID} \big),
\end{equation}
where $\phi$ can be instantiated as an MLP or a similarity-based function, quantifies the user’s predicted preference for the item.

\subsection{Model Optimization}
\subsubsection{Recommendation loss function}
Building upon the predicted preference scores, we employ the Bayesian Personalized Ranking (BPR) loss to guide the model toward ranking positive items higher than negative ones. For a user $u \in \mathcal{U}$ with a positive item $i \in \mathcal{I}^T$ such that $(u,i) \in \mathcal{S}^T_u$, and a negative item $i' \in \mathcal{I}^T$ such that $(u,i') \notin \mathcal{S}^T_u$, the loss is defined as
\begin{equation}
\mathcal{L}_{\text{rec}} = - \sum_{(u,i,i')} \log \sigma \big( \hat{y}_{u,i} - \hat{y}_{u,i'} \big),
\end{equation}
where $\hat{y}{u, i}$ and $\hat{y}{u, i'}$ are predicted scores for positive and negative items, and $\sigma(\cdot)$ is the sigmoid function. This loss encourages higher scores for observed target-domain interactions than for unobserved ones, aligning training with the ranking objective.

\subsubsection{Overall optimization objective}
The overall training objective combines the recommendation loss with the contrastive regularization terms, aiming to jointly optimize preference prediction and embedding consistency. Formally, the model minimizes
\begin{equation}
\mathcal{L} = \mathcal{L}_{\mathrm{rec}} + \lambda \, (\mathcal{L}_{\mathrm{c}}^1 + \, \mathcal{L}_{\mathrm{c}}^2),
\end{equation}
where $\lambda$ is the hyperparameter that balances the contrastive loss and recommendation loss. Optimizing this objective drives the model to integrate user interactions and item semantics across domains, providing a unified criterion for cross-domain recommendation.

\section{Experiment}

\subsection{Datasets and Experimental Setup}
\subsubsection{Datasets}
We conduct experiments on the Amazon Review Dataset~\footnote{http://jmcauley.ucsd.edu/data/amazon/}~\cite{mcauley2015image}, a widely used benchmark for cross-domain sequential recommendation tasks. Following the setup in~\cite{cao2022contrastive,ma2019pi}, we construct three bidirectional cross-domain scenarios: ``\textbf{Kitchen-Food}'' (comprising the \textit{Home and Kitchen} and \textit{Grocery and Gourmet Food} domains), ``\textbf{Movie-Book}'' (comprising the \textit{Movies and TV} and \textit{Books} domains), and ``\textbf{CDs-Movie}'' (comprising the \textit{CDs and Vinyl} and \textit{Movies and TV} domains). Consistent with standard CDR practices, these cross-domain scenarios include domains of varying relatedness, providing a comprehensive and diverse benchmark for cross-domain recommendation. Detailed data preprocessing and dataset statistics are summarized in Appendix~\ref{app:data}.

\begin{table*}[t!]
\centering
\setlength{\tabcolsep}{8pt} 
\setlength{\abovecaptionskip}{0.2mm}
\renewcommand{\arraystretch}{0.8}
\caption{Performance comparison results. ``(a)'' refers to single-domain sequential recommendation baselines, while ``(b)'' represents cross-domain sequential recommendation baselines. The best results are highlighted in bold, and the second-best results are underlined. ``RelaImpr'' reports the relative improvement over the strongest baseline.}
\label{tab:perf-kitchen-food}
\begin{tabular}{c l cccc c cccc}
\toprule
\multirow{2}{*}{\textbf{}} & \multirow{2}{*}{\textbf{Methods}} & \multicolumn{4}{c}{\textbf{Kitchen $\rightarrow$ Food}} & & \multicolumn{4}{c}{\textbf{Food $\rightarrow$ Kitchen}} \\
\cmidrule(lr){3-6} \cmidrule(lr){8-11}
 & & N@5 & N@10 & HR@5 & HR@10 & & N@5 & N@10 & HR@5 & HR@10 \\
\midrule
\multirow{2}{*}{(a)} & SASRec & 0.1754 & 0.2043 & 0.2348 & 0.3248 & & 0.1099 & 0.1356 & 0.1564 & 0.2369 \\
& CL4SRec & 0.1885 & 0.2261 & 0.2712 & 0.3879 & & 0.1176 & 0.1494 & 0.1750 & 0.2741 \\
\midrule
\multirow{6}{*}{(b)} & $\pi$-Net & 0.1653 & 0.1951 & 0.2258 & 0.3184 & & 0.1649 & 0.1944 & 0.2256 & 0.3172 \\
& C$^2$DSR & 0.1739 & 0.2056 & 0.2392 & 0.3328 & & 0.1381 & 0.1675 & 0.1968 & 0.2871 \\
& MGCL & 0.2191 & 0.2576 & 0.3142 & 0.4335 & & 0.1542 & 0.1902 & 0.2268 & 0.3389 \\
& CD-SASRec & 0.2147 & 0.2543 & 0.3071 & 0.4298 & & 0.1139 & 0.1418 & 0.1678 & 0.2547 \\
& Tri-CDR & 0.1892 & 0.2272 & 0.2746 & 0.3926 & & 0.1420 & 0.1797 & 0.2141 & 0.3315 \\
& TJAPL & \underline{0.2454} & \underline{0.2811} & \underline{0.3354} & \underline{0.4464} & & \underline{0.1719} & \underline{0.2084} & \underline{0.2486} & \underline{0.3616} \\
\midrule
\multirow{2}{*}{(Ours)} & \multirow{2}{*}{\textbf{SemaCDR}} & \textbf{0.2736} & \textbf{0.3105} & \textbf{0.3633} & \textbf{0.4775} & &\textbf{0.2047} & \textbf{0.2389} & \textbf{0.2744} & \textbf{0.3806} \\
&  & $\pm$ 0.0043 & $\pm$ 0.0050 & $\pm$ 0.0032 & $\pm$ 0.0023 & & $\pm$ 0.0035 & $\pm$ 0.0018 & $\pm$ 0.0035 & $\pm$ 0.0014 \\
\midrule
& \multicolumn{1}{l}{RelaImpr (\%)} & +11.5 & +10.5 & +8.3 & +7.0 & & +19.1 & +14.6 & +10.4 & +5.3 \\
\toprule
\multirow{2}{*}{\textbf{}} & \multirow{2}{*}{\textbf{Methods}} & \multicolumn{4}{c}{\textbf{Book $\rightarrow$ Movie}} & & \multicolumn{4}{c}{\textbf{Movie $\rightarrow$ Book}} \\
\cmidrule(lr){3-6} \cmidrule(lr){8-11}
 & & N@5 & N@10 & HR@5 & HR@10 & & N@5 & N@10 & HR@5 & HR@10 \\
\midrule
\multirow{2}{*}{(a)} & SASRec & 0.2246 & 0.2644 & 0.3117 & 0.4356 & & 0.1989 & 0.2345 & 0.2759 & 0.3864 \\
& CL4SRec & 0.2724 & 0.3077 & 0.3908 & 0.4909 & & 0.2442 & 0.2964 & 0.3609 & 0.4912 \\
\midrule
\multirow{6}{*}{(b)} & $\pi$-Net & 0.2171 & 0.2492 & 0.2924 & 0.3918 & & 0.2154 & 0.2475 & 0.2907 & 0.3905 \\
& C$^2$DSR & 0.2468 & 0.2815 & 0.3365 & 0.4469 & & 0.1748 & 0.2017 & 0.2577 & 0.3349 \\
& MGCL & 0.2382 & 0.2796 & 0.3469 & 0.4551 & & 0.2387 & 0.2788 & 0.3060 & 0.3996 \\
& CD-SASRec & 0.3254 & 0.3719 & 0.4568 & 0.6004 & & 0.3235 & 0.3716 & 0.4510 & 0.5998 \\
& Tri-CDR & 0.3258 & 0.3726 & 0.4548 & 0.5998 & & 0.2871 & 0.3328 & 0.4075 & 0.5488 \\
& TJAPL & \underline{0.3507} & \underline{0.3936} & \underline{0.4765} & \underline{0.6090} & & \underline{0.3434} & \underline{0.3844} & \underline{0.4701} & \underline{0.6019} \\
\midrule
\multirow{2}{*}{(Ours)} & \multirow{2}{*}{\textbf{SemaCDR}} & \textbf{0.3548} & \textbf{0.3992} & \textbf{0.4822} & \textbf{0.6199} & & \textbf{0.3607} & \textbf{0.4052} &  \textbf{0.4878} & \textbf{0.6256} \\
& & $\pm$ 0.0025 & $\pm$ 0.0015 & $\pm$ 0.0032 & $\pm$ 0.0070 & & $\pm$ 0.0041 & $\pm$ 0.0074 & $\pm$ 0.0040 & $\pm$ 0.0039 \\
\midrule
& \multicolumn{1}{l}{RelaImpr (\%)} & +1.2 & +1.4 & +1.2 & +1.8 & & +5.0 & +5.4 & +3.8 & +3.9 \\
\toprule
\multirow{2}{*}{\textbf{}} & \multirow{2}{*}{\textbf{Methods}} & \multicolumn{4}{c}{\textbf{CD $\rightarrow$ Movie}} & & \multicolumn{4}{c}{\textbf{Movie $\rightarrow$ CD}} \\
\cmidrule(lr){3-6} \cmidrule(lr){8-11}
 & & N@5 & N@10 & HR@5 & HR@10 & & N@5 & N@10 & HR@5 & HR@10 \\
\midrule
\multirow{2}{*}{(a)} & SASRec & 0.1631 & 0.1960 & 0.2269 & 0.3292 & & 0.1795 & 0.2131 & 0.2537 & 0.3584 \\
& CL4SRec & 0.1915 & 0.2370 & 0.2813 & 0.4224 & & 0.2243 & 0.2699 & 0.3307 & 0.4721 \\
\midrule
\multirow{6}{*}{(b)} & $\pi$-Net & 0.1434 & 0.1694 & 0.1912 & 0.2720 & & 0.1443 & 0.1709 & 0.1929 & 0.2758 \\
& C$^2$DSR & 0.1977 & 0.2287 & 0.2645 & 0.3620 & & 0.1808 & 0.2206 & 0.2476 & 0.3716 \\
& MGCL & 0.2308 & 0.2756 & 0.3313 & 0.4703 & & 0.2456 & 0.2940 & 0.3570 & 0.5074 \\
& CD-SASRec & 0.2306 & 0.2769 & 0.3288 & 0.4726 & & 0.2372 & 0.2778 & 0.3393 & 0.4651 \\
& Tri-CDR & 0.2322 & 0.2765 & 0.3371 & 0.4750 & & 0.2434 & 0.2926 & 0.3552 & 0.5084 \\
& TJAPL & \underline{0.2747} & \underline{0.3153} & \underline{0.3759} & \underline{0.5020} & & \underline{0.2869} & \underline{0.3278} & \underline{0.4018} & \underline{0.5287} \\
\midrule
\multirow{2}{*}{(Ours)} & \multirow{2}{*}{\textbf{SemaCDR}} & \textbf{0.3256} & \textbf{0.3725} & \textbf{0.4313} & \textbf{0.5760} & & \textbf{0.3185} & \textbf{0.3621} & \textbf{0.4215} & \textbf{0.5562} \\
& & $\pm$ 0.0081 & $\pm$ 0.0031 & $\pm$ 0.0047 & $\pm$ 0.0040 & & $\pm$ 0.0079 & $\pm$ 0.0071 & $\pm$ 0.0095 & $\pm$ 0.0075 \\
\midrule
& \multicolumn{1}{l}{RelaImpr (\%)} & +18.5 & +18.1 & +14.7 & +14.7 & & +11.0 &+10.5 & +4.9 & +5.2 \\
\bottomrule
\end{tabular}
\end{table*}
\subsubsection{Evaluation Metrics}
Following \cite{kang2018self,cao2022contrastive,ma2024triple}, we employ two standard metrics to evaluate top-N recommendation performance: Hit Ratio@k (\textbf{HR@k}) and Normalized Discounted Cumulative Gain@k (\textbf{NDCG@k}).
HR@k quantifies the proportion of test cases where the ground-truth item is present in the top-k recommendation list, while NDCG@k introduces positional awareness by assigning higher weights to relevant items that appear earlier in the ranked sequence. Our main results are reported in terms of HR@\{5, 10\} and NDCG@\{5, 10\}, where each value is averaged over five independent runs to ensure reliability and robustness.


\subsubsection{Implementation Details}
All experiments were implemented in PyTorch and trained on an NVIDIA RTX 3090 GPU with 24GB of memory. Baseline models were constructed following their officially released source codes. For a fair comparison, we adopt consistent training configurations across all methods, with the hidden dimension fixed at 64, batch size set to 256, dropout rate at 0.5, and the maximum sequence length truncated to 100. All model hyperparameters are tuned on the validation set to ensure optimal performance for both baselines and our SemaCDR model. During evaluation, we follow the standard practice in cross-domain recommendation \cite{xu2025multi,lin2020fissa}, sampling 100 negative items for each test case to balance efficiency and reliability. More implementation details about our model and baselines can be found in Appendix~\ref{app:implement}.

\begin{table*}[t!]
\centering 
\setlength{\abovecaptionskip}{0.2mm}
\setlength{\tabcolsep}{6pt} 
\renewcommand{\arraystretch}{0.8}
\caption{Compatibility analysis by integrating our proposed multi-view item semantics and contrastive learning loss across item views into two representative baselines, denoted as Enhanced C$^2$DSR (Tri-CDR).}
\label{tab:Compatibility experiment}
\begin{tabular}{l|| c| cccc|| c| cccc}
\toprule
\textbf{Methods} &  & N@5 & N@10 & HR@5 & HR@10 & & N@5 & N@10 & HR@5 & HR@10 \\
\midrule

C$^2$DSR 
& \multirow{3}{*}{\shortstack{\textbf{Kitchen}\\\\[-4pt]$\downarrow$\\\\[-4pt]\textbf{Food}}}
& 0.1739 & 0.2056 & 0.2392 & 0.3328 
& \multirow{3}{*}{\shortstack{\textbf{Food}\\\\[-4pt]$\downarrow$\\\\[-4pt]\textbf{Kitchen}}}
& 0.1381 & 0.1675 & 0.1968 & 0.2871 \\
Enhanced C$^2$DSR & & \textbf{0.1960} & \textbf{0.2240} & \textbf{0.2565} & \textbf{0.3570} & & \textbf{0.1558} & \textbf{0.1854} & \textbf{0.2169} & \textbf{0.3127} \\
RelaImpr (\%) & & +12.7 & +8.9 & +7.2 & +7.3 & & +12.8 & +10.7 & +10.2 & +9.0 \\
\cmidrule(lr){1-11}
Tri-CDR 
& \multirow{3}{*}{\shortstack{\textbf{Kitchen}\\\\[-4pt]$\downarrow$\\\\[-4pt]\textbf{Food}}}
& 0.1892 & 0.2272 & 0.2746 & 0.3926 
& \multirow{3}{*}{\shortstack{\textbf{Food}\\\\[-4pt]$\downarrow$\\\\[-4pt]\textbf{Kitchen}}}
& 0.1420 & 0.1797 & 0.2141 & 0.3315 \\
Enhanced Tri-CDR & & \textbf{0.1969} & \textbf{0.2348} & \textbf{0.2847} & \textbf{0.4106} & & \textbf{0.1469} & \textbf{0.1837} & \textbf{0.2162} & \textbf{0.3362} \\
RelaImpr (\%) & & +4.1 & +3.3 & +3.7 & +4.6 & & +3.5 & +2.2 & +1.0 & +1.4 \\
\midrule[\heavyrulewidth]

C$^2$DSR & \multirow{3}{*}{\shortstack{\textbf{Book}\\\\[-4pt]$\downarrow$\\\\[-4pt]\textbf{Movie}}} & 0.2468 & 0.2815 & 0.3365 & 0.4469 & \multirow{3}{*}{\shortstack{\textbf{Movie}\\\\[-4pt]$\downarrow$\\\\[-4pt]\textbf{Book}}} & 0.1748 & 0.2017 & 0.2577 & 0.3349 \\
Enhanced C$^2$DSR & & \textbf{0.2611} & \textbf{0.2965} & \textbf{0.3536} & \textbf{0.4637} & & \textbf{0.2011} & \textbf{0.2368} & \textbf{0.2809} & \textbf{0.3993} \\
RelaImpr (\%) & & +5.8 & +5.3 & +5.1 & +3.8 & & +15.1 & +17.4 & +9.0 & +19.2 \\
\cmidrule(lr){1-11}
Tri-CDR & \multirow{3}{*}{\shortstack{\textbf{Book}\\\\[-4pt]$\downarrow$\\\\[-4pt]\textbf{Movie}}} & 0.3258 & 0.3726 & 0.4548 & 0.5998 & \multirow{3}{*}{\shortstack{\textbf{Movie}\\\\[-4pt]$\downarrow$\\\\[-4pt]\textbf{Book}}} & 0.2871 & 0.3328 & 0.4075 & 0.5488 \\
Enhanced Tri-CDR & & \textbf{0.3306} & \textbf{0.3801} & \textbf{0.4607} & \textbf{0.6043} & & \textbf{0.2950} & \textbf{0.3420} & \textbf{0.4143} & \textbf{0.5601} \\
RelaImpr (\%) & & +1.5 & +2.0 & +1.3 & +0.8 & & +2.8 & +2.8 & +1.7 & +2.1 \\
\midrule[\heavyrulewidth]

C$^2$DSR & \multirow{3}{*}{\shortstack{\textbf{CD}\\\\[-4pt]$\downarrow$\\\\[-4pt]\textbf{Movie}}} & 0.1977 & 0.2287 & 0.2645 & 0.3620 & \multirow{3}{*}{\shortstack{\textbf{Movie}\\\\[-4pt]$\downarrow$\\\\[-4pt]\textbf{CD}}} & 0.1808 & 0.2206 & 0.2476 & 0.3716 \\
Enhanced C$^2$DSR & & \textbf{0.2103} & \textbf{0.2391} & \textbf{0.2764} & \textbf{0.3839} & & \textbf{0.1971} & \textbf{0.2312} & \textbf{0.2530} & \textbf{0.3827} \\
RelaImpr (\%) & & +6.4 & +4.5 & +4.5 & +6.1 & & +9.0 & +4.8 & +2.2 & +3.0 \\
\cmidrule(lr){1-11}
Tri-CDR & \multirow{3}{*}{\shortstack{\textbf{CD}\\\\[-4pt]$\downarrow$\\\\[-4pt]\textbf{Movie}}} & 0.2322 & 0.2765 & 0.3371 & 0.4750 & \multirow{3}{*}{\shortstack{\textbf{Movie}\\\\[-4pt]$\downarrow$\\\\[-4pt]\textbf{CD}}} & 0.2434 & 0.2926 & 0.3552 & 0.5084 \\
Enhanced Tri-CDR & & \textbf{0.2437} & \textbf{0.2832} & \textbf{0.3421} & \textbf{0.4829} & & \textbf{0.2513} & \textbf{0.3023} & \textbf{0.3679} & \textbf{0.5257} \\
RelaImpr (\%) & & +5.0 & +2.4 & +1.5 & +1.7 & & +3.2 & +3.3 & +3.6 & +3.4 \\
\bottomrule
\end{tabular}
\end{table*}
\subsection{Baselines}
We compare \model with several state-of-the-art methods to enable a comprehensive evaluation. The baselines cover two categories: \textit{(a) sequential recommendation models} that operate within a single domain, and \textit{(b) cross-domain sequential recommendation} approaches that directly target knowledge transfer across domains. This setup ensures that the evaluation reflects both the effectiveness of general sequence modeling techniques and the specific benefits of cross-domain modeling. For the sequential recommendation models, we select SASRec~\cite{kang2018self} and CL4SRec~\cite{xie2022contrastive} as baselines. For the cross-domain sequential recommendation models, we select $\pi$-Net~\cite{ma2019pi}, MGCL~\cite{xu2025multi}, C$^2$DSR~\cite{cao2022contrastive}, CD-SASRec~\cite{alharbi2021cross}, Tri-CDR~\cite{ma2024triple} and TJAPL~\cite{xu2024transfer} as baselines. More descriptions about baselines are summarized in Appendix~\ref{app:baseline}.

\subsection{Performance Comparison}
We compare \model with sequential recommendation models operating in a single domain and cross-domain approaches that leverage knowledge transfer. From the results in Table~\ref{tab:perf-kitchen-food}, several key observations can be drawn:

(1) Cross-domain recommendation methods generally achieve higher performance than single-domain sequential models, demonstrating that incorporating knowledge from related domains effectively enriches user representations and improves preference prediction. (2) Graph-based models (\eg MGCL and C$^2$DSR) generally outperform non-graph approaches (\eg $\pi$-Net and CD-SASRec), demonstrating that explicitly capturing associations among users and items enhances the representation of interaction patterns. Compared with relying solely on sequential item graphs, incorporating user–item connections like MGCL further enriches collaborative signal propagation and enables more effective preference modeling. (3) Explicitly modeling relationships between source- and target-domain sequences improves cross-domain knowledge transfer. Methods such as C$^2$DSR, TRI, and MGCL leverage inter-sequence designs, including mutual information maximization or contrastive learning, consistently outperforming models like $\pi$-Net, which highlights the benefit of incorporating sequence-level dependencies. (4) The inclusion of mixed-domain sequences in cross-domain baselines (\eg Tri-CDR) strengthens the model’s ability to capture user behavior by complementing domain-specific sequences, providing a more complete perspective and better modeling cross-domain temporal dependencies. (5) Our \model achieves consistently superior performance across all settings, demonstrating strong capability in cross-domain sequential recommendation. It integrates multi-view semantic embeddings with item-level contrastive learning, enhancing the alignment between semantic and collaborative signals and strengthening cross-domain generalization. Building on this, the bipartite graph captures rich user–item interactions for more effective relational modeling. Finally, cross-domain sequence fusion enhances alignment between source and target behaviors, while the inclusion of mixed-domain sequences further captures temporal dependencies. These components work together to deliver a coherent framework for preference prediction.








\subsection{Compatibility Analysis}
A key insight of this work is that item semantic features play a crucial role in cross-domain recommendation. To this end, we propose integrating multi-view item semantic embeddings that enrich item representations by combining ID embeddings with inner-domain semantic and domain-agnostic embeddings. Contrastive learning across these semantic views further aligns semantic and collaborative signals, enhancing representation expressiveness and transferability. This mechanism is general and can be readily integrated into existing cross-domain recommendation models. To assess its compatibility, we incorporate it into two representative baselines, C$^2$DSR and Tri-CDR. As shown in Table~\ref{tab:Compatibility experiment}, integrating our mechanism consistently improves performance for both methods, demonstrating its strong effectiveness. Notably, the two baselines represent distinct cross-domain modeling paradigms. One incorporates item-item graph structures while the other relies solely on sequential interactions. In both cases, integrating our mechanism yields substantial performance gains, highlighting that semantic enhancement consistently benefits cross-domain recommendation regardless of the underlying modeling approach.

\begin{figure}[htbp!]
    \centering
    \setlength{\abovecaptionskip}{0.2mm}
    \begin{subfigure}[b]{0.495\linewidth}
        \includegraphics[width=\linewidth]{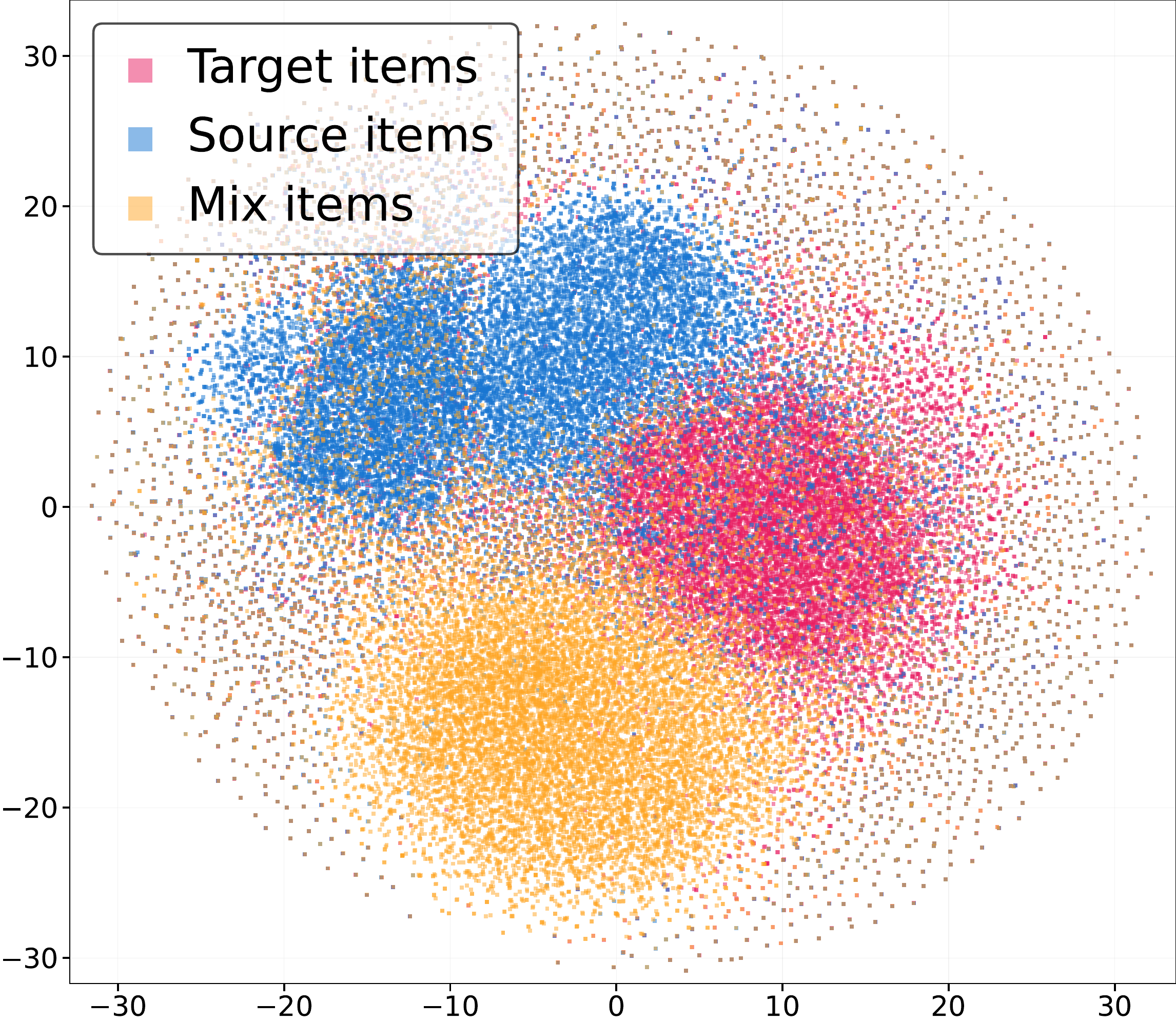}
        \caption{w/ DomAgnSem}
        \label{fig:item_emb_tsne_a}  
    \end{subfigure}
    \hfill  
    \begin{subfigure}[b]{0.495\linewidth}
        \includegraphics[width=\linewidth]{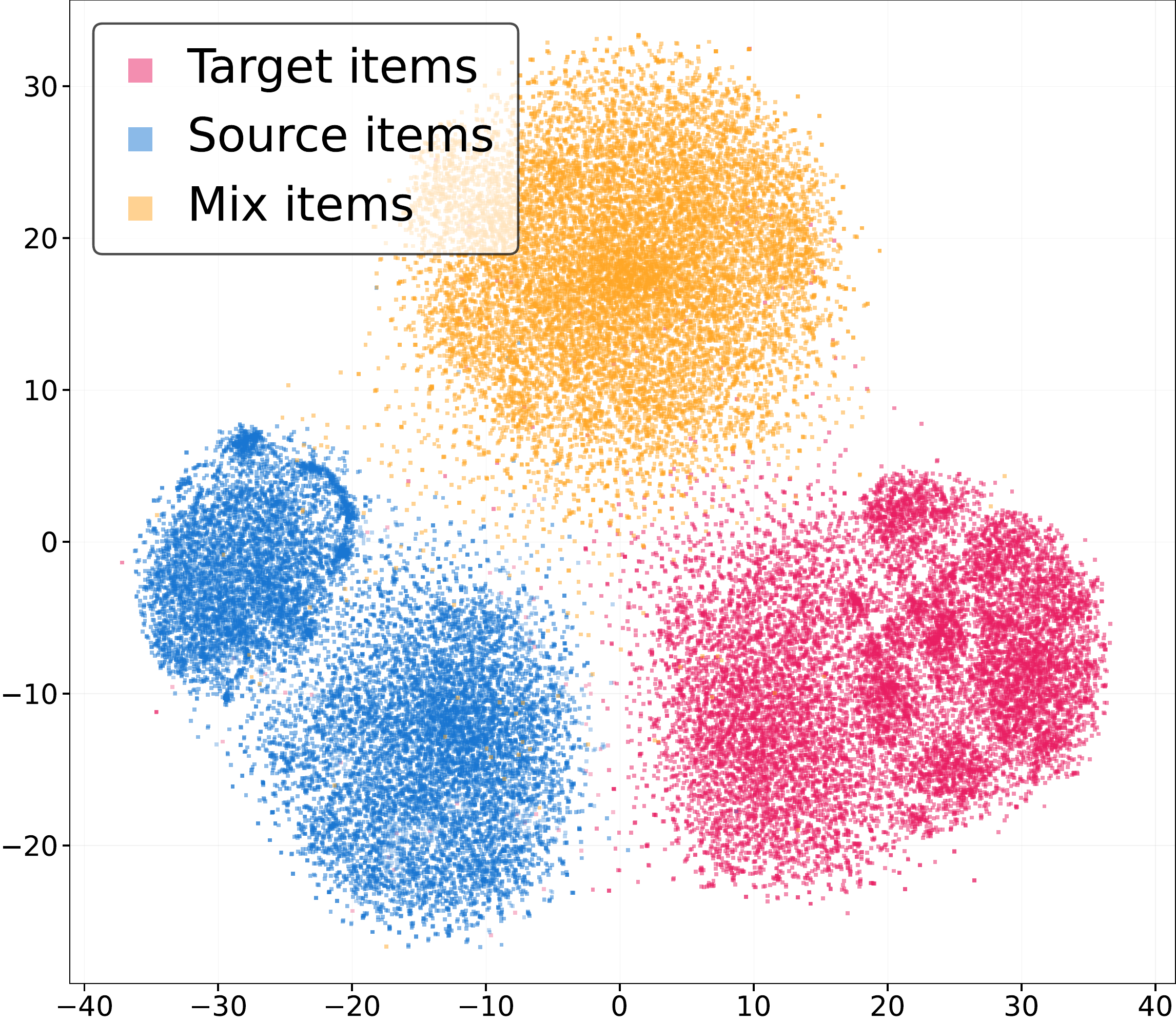}
        \caption{w/o DomAgnSem}
        \label{fig:item_emb_tsne_b}  
    \end{subfigure}
    \vspace{-3mm}  
    \caption{T-SNE distribution visualization of item embeddings learned from the graph-based collaborative learning module. ``DomAgnSem'' denotes domain-agnostic semantics.}
    \label{fig:item_emb_tsne}
\end{figure}

\subsection{Visualization}
Domain-agnostic semantic features are critical for establishing cross-domain associations and enabling knowledge transfer. To assess their effect, we conduct a visualization study on the item representations learned by the graph-based collaborative learning module on the Kitchen-Food dataset. Specifically, for the initial node embeddings, we compare two settings: one without domain-agnostic semantic embedding $\mathbf{e}_i = \mathbf{e}_i^{ID} \oplus \mathbf{e}_i^{inn}$, and one that additionally includes it $\mathbf{e}_i = \mathbf{e}_i^{ID} \oplus \mathbf{e}_i^{inn} \oplus \mathbf{e}_i^{agn}$. We project the item representations from the source domain, target domain, and mixed domain using t-SNE after GNN learning. As shown in Figure~\ref{fig:item_emb_tsne}, incorporating domain-agnostic semantics increases the overlap between source- and target-domain representations, indicating that these features effectively bridge the semantic gap across domains. These established associations support the subsequent integration of cross-domain sequences. When the representation spaces are closely aligned, the target domain can incorporate relevant signals from the source domain more effectively, facilitating smoother and more coherent cross-domain sequence fusion.

\subsection{Ablation Study}
To investigate the contribution of each key component in our SemaCDR framework, we perform ablation studies on the Kitchen–Food dataset (in the Kitchen-to-Food direction). Specifically, we design a set of model variants by removing or simplifying individual modules to assess their impact on overall performance. 
\begin{itemize}
    \item \textbf{w/o InnSem, w/o DomAgnSem, w/o InnDomAgnSem:} We examine the role of multi-view item semantics by individually excluding the inner-domain semantic, domain-agnostic, and both inner-domain and domain-agnostic embeddings from \model.
    \item \textbf{SimpGraph:} We simplify the graph-based collaborative learning module by retaining only item sequence graphs derived from each user’s independent interactions for the single domain, without constructing inter-item associations through users.
    \item \textbf{w/o CDBehav:} We disable the cross-domain behavior fusion mechanism, such that the source- and target-domain sequence embeddings are no longer merged together.
    \item \textbf{AvgFusion:} We replace the adaptive fusion with the averaging strategy, thereby removing the learnable fusion parameters.
    \item \textbf{w/o ConReg:} We remove the contrastive regularization between different item views to study its effect on embedding consistency.
\end{itemize}

%

The results in Table~\ref{tab:ablation_simple} clearly demonstrate the effectiveness of each component in our proposed \model. The full model consistently outperforms all its ablated variants across all metrics. The detailed observations are as follows: (1) Removing item semantic features causes the largest drop, indicating that domain-specific and domain-agnostic embeddings are complementary, and together they provide the essential semantic bridge for effective cross-domain recommendation. (2) Simplifying to single-domain item sequences discards inter-item associations captured via the full graph, demonstrating that leveraging cross-user connections is essential for richer collaborative information. (3) Integrating source- and target-domain sequences allows the model to exploit complementary dynamics from both domains, improving inter-domain alignment and enhancing target-domain representations relative to the version without fusion. (4) Compared with fixed fusion weights, our method learns adaptive fusion weights providing a flexible combination of information, yielding richer and more discriminative representations. (5) We observe that removing contrastive regularization between item views reduces performance, whereas our proposed contrastive regularization reinforces the alignment between domain-specific and domain-agnostic embeddings and maintains inter-item relational consistency, thereby improving the model’s ability to generalize user preferences.
\begin{table}[!t]
\centering
\setlength{\abovecaptionskip}{0.2mm}
\caption{Ablation study results on the Kitchen-Food dataset. Each row represents a variant of our model with a specific component removed or simplified.}
\label{tab:ablation_simple}
\begin{tabular}{lccccc}
\toprule
\textbf{Model Variant} & \textbf{N@5} & \textbf{N@10} & \textbf{H@5} & \textbf{H@10} \\
\midrule
w/o InnSem & 0.2608 & 0.2993 & 0.3459 & 0.4637 \\
w/o DomAgnSem & 0.2502 & 0.2896 & 0.3392 & 0.4591 \\
w/o InnDomAgnSem & 0.2370 & 0.2740 & 0.3258 & 0.4404 \\
SimpGraph & 0.2698 & 0.3014 & 0.3505 & 0.4721 \\
w/o CDBehav & 0.2671 & 0.3009 & 0.3469 & 0.4678 \\
AvgFusion & 0.2510 & 0.2867 & 0.3373 & 0.4479 \\
w/o ConReg & 0.2633 & 0.2948 & 0.3510 & 0.4631 \\
\midrule
\textbf{SemaCDR (Full)} & \textbf{0.2736} & \textbf{0.3105} & \textbf{0.3633} & \textbf{0.4775} \\
\bottomrule
\end{tabular}
\end{table} 



\subsection{Hyper-parameter Analysis}
We conduct a hyperparameter study on the Kitchen–Food dataset (HR@10) to validate the robustness of SemaCDR. The analysis shows that setting the number of domain-agnostic semantic categories to $K=8$ achieves the best trade-off between expressiveness and redundancy. For the contrastive regularization weight, performance improves as $\lambda$ increases and peaks at $\lambda=0.1$, beyond which the main recommendation objective is harmed. In addition, using a smaller learning rate for semantic-related modules ($3e^{-4}$) and a larger learning rate for general components ($1e^{-3}$) yields the best results, aligning with the stability of semantic representations. Detailed results are provided in the Appendix~\ref{app:hyper}.

\section{Related Work}
\subsection{LLM-based Recommender Systems}
LLM-based recommender systems have emerged to address challenges such as data sparsity and cold-start, leveraging their strengths in semantic understanding, reasoning, and zero/few-shot learning~\cite{liu2024large,wu2024survey,chen2024large,zhao2024recommender}. Existing work can be broadly categorized into two paradigms: LLM-Enhanced Recommender Systems (LLMERS)~\cite{xi2024towards,wang2024llmrg,liu2025llmemb,huang2025large} and LLM-as-Recommender systems~\cite{liang2024taxonomy,petruzzelli2024instructing,bao2023tallrec}. \textbf{LLMERS} uses LLMs to assist conventional recommendation models, which remain the primary engines. Representative approaches include enriching inputs by extracting user preferences and item semantics~\cite{xi2024towards,wang2024llmrg}, enhancing representations with LLM-generated embeddings~\cite{liu2025llmemb}, and refining ranking or re-ranking processes via LLM reasoning~\cite{huang2025large}. These methods demonstrate how LLMs augment traditional recommenders with richer semantics and more flexible reasoning. In contrast, \textbf{LLM-as-Recommender} treats LLMs as the core recommendation engines, either through prompt-based inference without additional training~\cite{liang2024taxonomy} or via task-specific fine-tuning~\cite{petruzzelli2024instructing,bao2023tallrec}. This paradigm highlights the potential of LLMs to function as standalone recommenders, offering flexibility but also raising new challenges in efficiency and scalability.

\subsection{Cross Domain Recommendations}
Cross-Domain Recommendation (CDR) has gained increasing attention as a means of transferring knowledge from data-rich source domains to improve recommendations in target domains~\cite{zang2022survey,zhang2025comprehensive,zhu2021cross,zhang2024m3oe}. Existing studies can be broadly categorized into three methodological streams. The first investigates \textit{mapping-based approaches} that transfer knowledge across domains by aligning the latent embeddings of users or items. These methods generate domain-specific embeddings and then learn mappings or impose shared constraints to bridge the source and target embedding spaces, facilitating effective cross-domain knowledge transfer~\cite{man2017cross,ma2019pi,chen2025sustainability}. The second category, \textit{fusion-based approaches}, seeks to model user preferences by jointly leveraging information from multiple domains. These methods capture cross-domain interactions and align representations across domains to enhance the robustness and expressiveness of user and item modeling, thereby effectively addressing domain shifts and mitigating negative transfer~\cite{xu2025multi,ma2024triple,cao2022contrastive,alharbi2021cross,xu2024transfer,tao2025task}. The third category explores \textit{the integration of LLMs into cross-domain recommendation}, where LLMs are incorporated as part of the recommendation model to enhance representation learning or to act as adaptable recommendation engines, thereby facilitating knowledge transfer and improving performance across domains~\cite{petruzzelli2024instructing}. Despite these advances, existing CDR methods rely heavily on domain-specific features or embeddings, limiting their ability to capture transferable semantics across domains. In this paper, we propose leveraging LLMs to construct a unified semantic space, directly modeling high-level, domain-agnostic representations to facilitate more effective cross-domain knowledge transfer.

\section{Conclusion}
In this work, we present \model, a semantics-driven framework for cross-domain sequential recommendation that leverages large language models to construct a unified semantic space. By integrating domain-specific and domain-agnostic item representations, \model effectively captures multi-dimensional user preferences and aligns interaction sequences across source and target domains through a cross-domain behavior fusion mechanism. The adaptive fusion prediction module further combines sequential, user, and mixed-domain representations, producing cohesive preference estimations. Comprehensive evaluations on real-world datasets demonstrate that \model consistently outperforms existing state-of-the-art methods, validating the benefits of unified semantic modeling and cross-domain alignment. Our findings highlight the potential of LLMs as powerful tools for capturing transferable knowledge, offering a promising direction for developing more robust and generalizable cross-domain recommender systems.

\section{Acknowledgments}
Chunxu Zhang and Bo Yang are supported by the National Natural Science Foundation of China under Grant Nos. U22A2098, 62172185, 62206105 and 62202200; the Major Science and Technology Development Plan of Jilin Province under Grant No.20240212003GX, the Major Science and Technology Development Plan of Changchun under Grant No.2024WX05. Zijian Zhang is supported by the China Postdoctoral Science Foundation (2025M771587), the Open Funding Programs of State Key Laboratory of AI Safety and the Key R\&D Program of Jilin Province (20260205050GH). Irwin King is supported by the Research Grants Council of the Hong Kong Special Administrative Region, China (CUHK 2410072, RGC R1015-23) and (CUHK 2300246, RGC C1043-24G).



\bibliographystyle{ACM-Reference-Format}
\bibliography{sample-base}

@String{Computing = "Computing" }

@String{Springer = "Springer-Verlag" }

@ArtifactSoftware{R,
    title = {R: A Language and Environment for Statistical Computing},
    author = {{R Core Team}},
    organization = {R Foundation for Statistical Computing},
    address = {Vienna, Austria},
    year = {2019},
    url = {https://www.R-project.org/},
}

@inproceedings{man2017cross,
  title={Cross-domain recommendation: An embedding and mapping approach.},
  author={Man, Tong and Shen, Huawei and Jin, Xiaolong and Cheng, Xueqi},
  booktitle={Ijcai},
  volume={17},
  pages={2464--2470},
  year={2017}
}

@inproceedings{zhu2021transfer,
  title={Transfer-meta framework for cross-domain recommendation to cold-start users},
  author={Zhu, Yongchun and Ge, Kaikai and Zhuang, Fuzhen and Xie, Ruobing and Xi, Dongbo and Zhang, Xu and Lin, Leyu and He, Qing},
  booktitle={Proceedings of the 44th international ACM SIGIR conference on research and development in information retrieval},
  pages={1813--1817},
  year={2021}
}

@article{li2023preference,
  title={Preference-aware graph attention networks for cross-domain recommendations with collaborative knowledge graph},
  author={Li, Yakun and Hou, Lei and Li, Juanzi},
  journal={ACM transactions on information systems},
  volume={41},
  number={3},
  pages={1--26},
  year={2023},
  publisher={ACM New York, NY}
}

@article{guo2021gcn,
  title={DA-GCN: A domain-aware attentive graph convolution network for shared-account cross-domain sequential recommendation},
  author={Guo, Lei and Tang, Li and Chen, Tong and Zhu, Lei and Nguyen, Quoc Viet Hung and Yin, Hongzhi},
  journal={arXiv preprint arXiv:2105.03300},
  year={2021}
}

@inproceedings{cao2023towards,
  title={Towards universal cross-domain recommendation},
  author={Cao, Jiangxia and Li, Shaoshuai and Yu, Bowen and Guo, Xiaobo and Liu, Tingwen and Wang, Bin},
  booktitle={Proceedings of the Sixteenth ACM International Conference on web search and data mining},
  pages={78--86},
  year={2023}
}

@inproceedings{mcauley2015image,
  title={Image-based recommendations on styles and substitutes},
  author={McAuley, Julian and Targett, Christopher and Shi, Qinfeng and Van Den Hengel, Anton},
  booktitle={Proceedings of the 38th international ACM SIGIR conference on research and development in information retrieval},
  pages={43--52},
  year={2015}
}

@inproceedings{cao2022contrastive,
  title={Contrastive cross-domain sequential recommendation},
  author={Cao, Jiangxia and Cong, Xin and Sheng, Jiawei and Liu, Tingwen and Wang, Bin},
  booktitle={Proceedings of the 31st ACM International Conference on Information \& Knowledge Management},
  pages={138--147},
  year={2022}
}

@article{xu2025multi,
  title={A multi-view graph contrastive learning framework for cross-domain sequential recommendation},
  author={Xu, Zitao and Chen, Shu and Pan, Weike and Ming, Zhong},
  journal={ACM Transactions on Recommender Systems},
  volume={3},
  number={4},
  pages={1--28},
  year={2025},
  publisher={ACM New York, NY}
}

@inproceedings{xu2025heterogeneous,
  title={Heterogeneous Graph Transfer Learning for Category-aware Cross-Domain Sequential Recommendation},
  author={Xu, Zitao and Chen, Xiaoqing and Pan, Weike and Ming, Zhong},
  booktitle={Proceedings of the ACM on Web Conference 2025},
  pages={1951--1962},
  year={2025}
}

@inproceedings{ma2019pi,
  title={$\pi$-net: A parallel information-sharing network for shared-account cross-domain sequential recommendations},
  author={Ma, Muyang and Ren, Pengjie and Lin, Yujie and Chen, Zhumin and Ma, Jun and Rijke, Maarten de},
  booktitle={Proceedings of the 42nd international ACM SIGIR conference on research and development in information retrieval},
  pages={685--694},
  year={2019}
}

@inproceedings{kang2018self,
  title={Self-attentive sequential recommendation},
  author={Kang, Wang-Cheng and McAuley, Julian},
  booktitle={2018 IEEE international conference on data mining (ICDM)},
  pages={197--206},
  year={2018},
  organization={IEEE}
}

@inproceedings{lin2020fissa,
  title={FISSA: Fusing item similarity models with self-attention networks for sequential recommendation},
  author={Lin, Jing and Pan, Weike and Ming, Zhong},
  booktitle={Proceedings of the 14th ACM conference on recommender systems},
  pages={130--139},
  year={2020}
}

@article{ma2024triple,
  title={Triple sequence learning for cross-domain recommendation},
  author={Ma, Haokai and Xie, Ruobing and Meng, Lei and Chen, Xin and Zhang, Xu and Lin, Leyu and Zhou, Jie},
  journal={ACM Transactions on Information Systems},
  volume={42},
  number={4},
  pages={1--29},
  year={2024},
  publisher={ACM New York, NY}
}

@article{lyu2023llm,
  title={Llm-rec: Personalized recommendation via prompting large language models},
  author={Lyu, Hanjia and Jiang, Song and Zeng, Hanqing and Xia, Yinglong and Wang, Qifan and Zhang, Si and Chen, Ren and Leung, Christopher and Tang, Jiajie and Luo, Jiebo},
  journal={arXiv preprint arXiv:2307.15780},
  year={2023}
}

@article{wang2024towards,
  title={Towards next-generation llm-based recommender systems: A survey and beyond},
  author={Wang, Qi and Li, Jindong and Wang, Shiqi and Xing, Qianli and Niu, Runliang and Kong, He and Li, Rui and Long, Guodong and Chang, Yi and Zhang, Chengqi},
  journal={arXiv preprint arXiv:2410.19744},
  year={2024}
}

@article{liu2024large,
  title={Large Language Model Enhanced Recommender Systems: A Survey},
  author={Liu, Qidong and Zhao, Xiangyu and Wang, Yuhao and Wang, Yejing and Zhang, Zijian and Sun, Yuqi and Li, Xiang and Wang, Maolin and Jia, Pengyue and Chen, Chong and others},
  journal={arXiv preprint arXiv:2412.13432},
  year={2024}
}

@article{wu2024survey,
  title={A survey on large language models for recommendation},
  author={Wu, Likang and Zheng, Zhi and Qiu, Zhaopeng and Wang, Hao and Gu, Hongchao and Shen, Tingjia and Qin, Chuan and Zhu, Chen and Zhu, Hengshu and Liu, Qi and others},
  journal={World Wide Web},
  volume={27},
  number={5},
  pages={60},
  year={2024},
  publisher={Springer}
}

@article{chen2024large,
  title={When large language models meet personalization: Perspectives of challenges and opportunities},
  author={Chen, Jin and Liu, Zheng and Huang, Xu and Wu, Chenwang and Liu, Qi and Jiang, Gangwei and Pu, Yuanhao and Lei, Yuxuan and Chen, Xiaolong and Wang, Xingmei and others},
  journal={World Wide Web},
  volume={27},
  number={4},
  pages={42},
  year={2024},
  publisher={Springer}
}

@article{zhao2024recommender,
  title={Recommender systems in the era of large language models (llms)},
  author={Zhao, Zihuai and Fan, Wenqi and Li, Jiatong and Liu, Yunqing and Mei, Xiaowei and Wang, Yiqi and Wen, Zhen and Wang, Fei and Zhao, Xiangyu and Tang, Jiliang and others},
  journal={IEEE Transactions on Knowledge and Data Engineering},
  volume={36},
  number={11},
  pages={6889--6907},
  year={2024},
  publisher={IEEE}
}

@inproceedings{xi2024towards,
  title={Towards open-world recommendation with knowledge augmentation from large language models},
  author={Xi, Yunjia and Liu, Weiwen and Lin, Jianghao and Cai, Xiaoling and Zhu, Hong and Zhu, Jieming and Chen, Bo and Tang, Ruiming and Zhang, Weinan and Yu, Yong},
  booktitle={Proceedings of the 18th ACM Conference on Recommender Systems},
  pages={12--22},
  year={2024}
}

@inproceedings{wang2024llmrg,
  title={Llmrg: Improving recommendations through large language model reasoning graphs},
  author={Wang, Yan and Chu, Zhixuan and Ouyang, Xin and Wang, Simeng and Hao, Hongyan and Shen, Yue and Gu, Jinjie and Xue, Siqiao and Zhang, James and Cui, Qing and others},
  booktitle={Proceedings of the AAAI conference on artificial intelligence},
  volume={38},
  number={17},
  pages={19189--19196},
  year={2024}
}

@inproceedings{huang2025large,
  title={Large Language Model Simulator for Cold-Start Recommendation},
  author={Huang, Feiran and Bei, Yuanchen and Yang, Zhenghang and Jiang, Junyi and Chen, Hao and Shen, Qijie and Wang, Senzhang and Karray, Fakhri and Yu, Philip S},
  booktitle={Proceedings of the Eighteenth ACM International Conference on Web Search and Data Mining},
  pages={261--270},
  year={2025}
}

@inproceedings{liu2025llmemb,
  title={Llmemb: Large language model can be a good embedding generator for sequential recommendation},
  author={Liu, Qidong and Wu, Xian and Wang, Wanyu and Wang, Yejing and Zhu, Yuanshao and Zhao, Xiangyu and Tian, Feng and Zheng, Yefeng},
  booktitle={Proceedings of the AAAI Conference on Artificial Intelligence},
  volume={39},
  number={11},
  pages={12183--12191},
  year={2025}
}

@article{liang2024taxonomy,
  title={Taxonomy-guided zero-shot recommendations with llms},
  author={Liang, Yueqing and Yang, Liangwei and Wang, Chen and Xu, Xiongxiao and Yu, Philip S and Shu, Kai},
  journal={arXiv preprint arXiv:2406.14043},
  year={2024}
}

@inproceedings{petruzzelli2024instructing,
  title={Instructing and prompting large language models for explainable cross-domain recommendations},
  author={Petruzzelli, Alessandro and Musto, Cataldo and Laraspata, Lucrezia and Rinaldi, Ivan and de Gemmis, Marco and Lops, Pasquale and Semeraro, Giovanni},
  booktitle={Proceedings of the 18th ACM Conference on Recommender Systems},
  pages={298--308},
  year={2024}
}

@inproceedings{bao2023tallrec,
  title={Tallrec: An effective and efficient tuning framework to align large language model with recommendation},
  author={Bao, Keqin and Zhang, Jizhi and Zhang, Yang and Wang, Wenjie and Feng, Fuli and He, Xiangnan},
  booktitle={Proceedings of the 17th ACM conference on recommender systems},
  pages={1007--1014},
  year={2023}
}

@article{zang2022survey,
  title={A survey on cross-domain recommendation: taxonomies, methods, and future directions},
  author={Zang, Tianzi and Zhu, Yanmin and Liu, Haobing and Zhang, Ruohan and Yu, Jiadi},
  journal={ACM Transactions on Information Systems},
  volume={41},
  number={2},
  pages={1--39},
  year={2022},
  publisher={ACM New York, NY}
}

@inproceedings{park2024pacer,
  title={Pacer and runner: Cooperative learning framework between single-and cross-domain sequential recommendation},
  author={Park, Chung and Kim, Taesan and Yoon, Hyungjun and Hong, Junui and Yu, Yelim and Cho, Mincheol and Choi, Minsung and Choo, Jaegul},
  booktitle={Proceedings of the 47th International ACM SIGIR Conference on Research and Development in Information Retrieval},
  pages={2071--2080},
  year={2024}
}

@article{zhang2025comprehensive,
  title={A comprehensive survey on cross-domain recommendation: Taxonomy, progress, and prospects},
  author={Zhang, Hao and Cheng, Mingyue and Liu, Qi and Jiang, Junzhe and Wang, Xianquan and Zhang, Rujiao and Lei, Chenyi and Chen, Enhong},
  journal={arXiv preprint arXiv:2503.14110},
  year={2025}
}

@article{zhu2021cross,
  title={Cross-domain recommendation: challenges, progress, and prospects},
  author={Zhu, Feng and Wang, Yan and Chen, Chaochao and Zhou, Jun and Li, Longfei and Liu, Guanfeng},
  journal={arXiv preprint arXiv:2103.01696},
  year={2021}
}

@inproceedings{alharbi2021cross,
  title={Cross-domain self-attentive sequential recommendations},
  author={Alharbi, Nawaf and Caragea, Doina},
  booktitle={Proceedings of International Conference on Data Science and Applications: ICDSA 2021, Volume 2},
  pages={601--614},
  year={2021},
  organization={Springer}
}

@article{xu2024transfer,
  title={Transfer learning in cross-domain sequential recommendation},
  author={Xu, Zitao and Pan, Weike and Ming, Zhong},
  journal={Information Sciences},
  volume={669},
  pages={120550},
  year={2024},
  publisher={Elsevier}
}

@inproceedings{xie2022contrastive,
  title={Contrastive learning for sequential recommendation},
  author={Xie, Xu and Sun, Fei and Liu, Zhaoyang and Wu, Shiwen and Gao, Jinyang and Zhang, Jiandong and Ding, Bolin and Cui, Bin},
  booktitle={2022 IEEE 38th international conference on data engineering (ICDE)},
  pages={1259--1273},
  year={2022},
  organization={IEEE}
}

@inproceedings{liu2025spottrip,
  title={{SPOT-Trip: Dual-Preference Driven Out-of-Town Trip Recommendation}},
  author={Liu, Yinghui and Miao, Hao and Shen, Guojiang and Zhao, Yan and Kong, Xiangjie and Lee, Ivan},
  booktitle={{NeurIPS}},
  year={2025}
}

@inproceedings{chen2025sustainability,
  title={Sustainability-Oriented Task Recommendation in Spatial Crowdsourcing},
  author={Chen, Jinwen and Miao, Hao and Qiu, Dazhuo and Guo, Jiannan and Li, Yawen and Zhao, Yan},
  booktitle={ICDE},
  pages={2712--2725},
  year={2025}
}

@article{tao2025task,
  title={Task-Aware Retrieval Augmentation for Dynamic Recommendation},
  author={Tao, Zhen and Jiang, Xinke and Feng, Qingshuai and Zhang, Haoyu and Du, Lun and Fang, Yuchen and Miao, Hao and Xie, Bangquan and Sun, Qingqiang},
  journal={arXiv preprint arXiv:2511.12495},
  year={2025}
}

@inproceedings{zhang2024m3oe,
  title={M3oe: Multi-domain multi-task mixture-of experts recommendation framework},
  author={Zhang, Zijian and Liu, Shuchang and Yu, Jiaao and Cai, Qingpeng and Zhao, Xiangyu and Zhang, Chunxu and Liu, Ziru and Liu, Qidong and Zhao, Hongwei and Hu, Lantao and others},
  booktitle={Proceedings of the 47th International ACM SIGIR Conference on Research and Development in Information Retrieval},
  pages={893--902},
  year={2024}
}

\appendix

\section{LLM-generated Features Illustration}\label{app:llm}
In this section, we present an illustrative example demonstrating how domain-agnostic semantic categories and enriched inner-domain semantics are generated using LLMs. Using the Movie–Book CDR scenario as a case study, we design tailored prompts that guide the LLM to produce semantically coherent categories shared across domains, along with detailed inner-domain semantics. For clarity, we provide both the prompt templates and representative generation results.
Here is the prompt template $T_{rec}$ and the corresponding response with a sample:
\begin{tcolorbox}[%
    colback=gray!10,
    colframe=gray,
    width=1\linewidth,
    arc=1mm, 
    auto outer arc,
    title={Cross-Domain Recommendation Prompt Template $T_{rec}$},
    breakable,]
    
    You are a professional cross-domain recommendation expert. I will provide you with titles/descriptions for items that can be either \textbf{Books} or \textbf{Movies and TV}.
    
    Given an item with the category \textcolor{blue}{[Books; Teen \& Science Fiction \& Fantasy]}, brand \textcolor{blue}{[Visit Amazon's Lloyd Alexander Page]}, and the following title/description: \textcolor{blue}{[The Prydain Chronicles]}.
    
    Your task is to perform two main actions based on the input text and your world knowledge:
    \begin{enumerate}[noitemsep, topsep=2pt] 
        \item  \textbf{Extract Domain-Agnostic Features:} Using the "Classification Schema" provided below, extract relevant domain-agnostic features. For EACH category in the schema, you MUST select at least one relevant label.
        \item  \textbf{Generate a Semantic Summary:} Create a single, concise sentence that serves as a semantic summary. This summary should encapsulate the essence of the item and bridge commonalities that could appeal to users interested in items with similar characteristics, regardless of their domain.
    \end{enumerate}
    
    \textbf{Classification Schema:}
    \begin{verbatim}
[
{"Genre": ["Action", "Comedy", "Drama", 
  "Thriller", "Sci-Fi", "Romance"...]},
  {"Target Audience": ["Children", "Young Adult", 
  "Adult"...]},
  {"Themes": ["Love", "Loss", "Friendship", 
  "Betrayal", "Revenge"... ]},
  ...
]
    \end{verbatim}
    
    \textbf{Conversion Rules:}
    \begin{enumerate}[noitemsep, topsep=2pt]
        \item Analyze the input title/description. Leverage your world knowledge.
        \item For EACH category in the "Classification Schema", you MUST select one or more labels.
        \item Generate the "SemanticSummary" sentence as described above.
        \item Output strictly in the following JSON format, including all categories:
    \end{enumerate}
    
    \begin{tcolorbox}[colback=white, colframe=black!10, boxrule=0.5pt, fontupper=\ttfamily]
    \{ \\
    \quad "Features": \{ \\
    \quad\quad  "Genre": [\textcolor{red}{\textless list of genres\textgreater}], \\
    \quad\quad "Target Audience": [\textcolor{red}{\textless list of audiences\textgreater}], \\
    \quad\quad "Themes": [\textcolor{red}{\textless list of themes\textgreater}], \\
    \quad\quad ... \\
    \quad \}, \\
    \quad "SemanticSummary": "\textcolor{red}{a single, concise sentence summarizing...}" \\
    \}
    \end{tcolorbox}
\end{tcolorbox}

\begin{tcolorbox}[
    colframe=gray,
    width=\linewidth,
    arc=1mm,
    auto outer arc,
    breakable,
    title={LLM Response Example},
    colback=gray!10,
    colbacklower=white,
    fontlower=\ttfamily
]

\{ \\
\quad "mapped\_id": "708", \\
\quad "features": \{ \\
\quad\quad "Genre": ["Drama", "Historical"], \\
\quad\quad "Target Audience": ["Young Adult"], \\
\quad\quad "Themes": ["Social Issues", "Political Intrigue"], \\
\quad\quad "Setting": ["Historical", "Urban"], \\
\quad\quad "Tone": ["Serious", "Dark"], \\
\quad\quad "Emotional Arc": ["Stressful"], \\
\quad\quad "Pace": ["Moderate-Paced"], \\
\quad\quad "Format/Length": ["Medium"], \\
\quad\quad "Origin": ["Original Work"], \\
\quad\quad "Narrative Style": ["Third Person Limited"] \\
\quad \}, \\
\quad "semantic\_summary": "This book, likely aimed at young adults, offers a serious exploration of social and political themes within a historical setting, potentially appealing to readers interested in thought-provoking, character-driven narratives." \\
\}
\end{tcolorbox}

To illustrate the characteristics of LLM-generated domain-agnostic semantics, we further examine their distributional patterns at the user level. Specifically, we randomly select a user from the dataset Food–Kitchen, and identify the domain-agnostic categories and subcategories of the items they interacted with in both the source and target domains. We then compute the frequency of each category (subcategory) and visualize the results using histograms. As shown in Figure~\ref{model}, the user exhibits highly similar semantic distributions across the two domains, which aligns with the intuitive expectation that user preferences tend to be stable and transferable. For example, a user who favors ``Historical'' or ``Drama'' themes in movies is likely to prefer books of similar genres. This observation confirms that the extracted domain-agnostic semantics effectively capture stable, transferable user interests, aligning well with the core objective of cross-domain recommendation.

\section{Data Preprocessing and Dataset Statistics}\label{app:data}
Based on prior work~\cite{kang2018self,xu2025heterogeneous,xu2025multi}, we preprocess the data in the following steps: (i) treat all interactions as positive feedback and order them by timestamp to construct user sequences; (ii) retain items with at least 10 interactions and users with at least 3 interactions per domain, while removing duplicates; (iii) keep only users that appear in both domains of each pair; and (iv) apply the leave-one-out splitting strategy, where for each user, the last interaction is used for testing, the second-to-last for validation, and the remainder for training. The dataset statistics are summarized in Table~\ref{tab:dataset_stats}.
\begin{table}[t!]
\setlength{\textfloatsep}{5pt} 
\centering
\renewcommand{\arraystretch}{0.8}
\caption{Statistics of three cross-domain datasets.}
\setlength{\belowcaptionskip}{2pt}
\label{tab:dataset_stats}
\normalsize 
\setlength{\tabcolsep}{3.5pt} 
\begin{tabular}{lcccc}
\toprule
Dataset & Overlapped Users & \# Items & \# Interactions & Avg. Len. \\
\midrule
Movie     & 8,537      & 10,211   & 268,413   & 31.44     \\
CD        & 8,537      & 7,450    & 153,000   & 17.92     \\
\midrule
Book      & 30,938     & 31,067   & 664,205   & 21.47     \\
Movie     & 30,938     & 19,975   & 763,301   & 24.67     \\
\midrule
Food      & 19,229     & 8,102    & 232,717   & 12.10     \\
Kitchen   & 19,229     & 12,106   & 274,842   & 14.29     \\
\bottomrule
\end{tabular}
\end{table}

\section{More Implementation Details}\label{app:implement}
In our method, the graph-based collaborative learning module utilizes a model based on Graph Convolutional Network (GCN). Each GCN layer transforms normalized neighborhood features via a learnable weight matrix and bias, followed by LeakyReLU activation. This GCN design achieves a good balance between effectiveness and efficiency, requiring only one layer of propagation operation to complete. For the multi-view sequence representation learning module, the sequence encoder adopts a Transformer architecture with single-head attention and two attention blocks (\ie \( \text{num\_heads}=1 \), \( \text{num\_blocks}=2 \)). Contrastive learning utilizes the InfoNCE loss function, with a temperature parameter of \( \tau=0.07 \) and a contrastive loss weight of 0.1. For LLM-based semantic extraction, we adopt a zero-shot setting without fine-tuning. Outliers are filtered via feature word frequency, and non-conforming outputs are re-generated to ensure semantic quality. 

For baseline methods using sequence encoders~\cite{kang2018self, alharbi2021cross, xu2025multi, xie2022contrastive, ma2024triple, xu2024transfer, cao2022contrastive}, we uniformly retain the single-head attention mechanism and dual-attention-layer configuration to ensure a fair comparison. For all GNN-based methods~\cite{xu2025multi,cao2022contrastive}, we assume a layer depth of 1 and follow the recommended parameters provided by the original authors. For the single-domain recommendation model CL4Rec, \( \text{cl\_embs} \) is set to the "predict mode", \( \text{w\_clloss}=0.1 \), and other parameters use the default settings provided by the original authors. For certain models that support multi-source domain recommendations~\cite{xu2025multi, xu2024transfer}, only one domain is designated as the source domain to ensure fairness.

We further conducted compatibility experiments for this model framework. The hyperparameter settings for the newly added semantic relevance module in different models are detailed as follows: For the Tri-CDR model~\cite{ma2024triple}: Learning rate = 0.0005, contrastive loss weight = 1, temperature parameter \( \tau=0.2 \), maximum sequence length \( L=100 \). For the C$^2$DSR model~\cite{cao2022contrastive}: Learning rate = 0.0001, temperature parameter \( \tau=0.07 \), maximum sequence length \( L=100 \), contrastive loss weight = 0.15. All other parameters retain the default settings.

\begin{figure}[!t]
\setlength{\belowcaptionskip}{-2mm}
{{\includegraphics[width=1\linewidth]{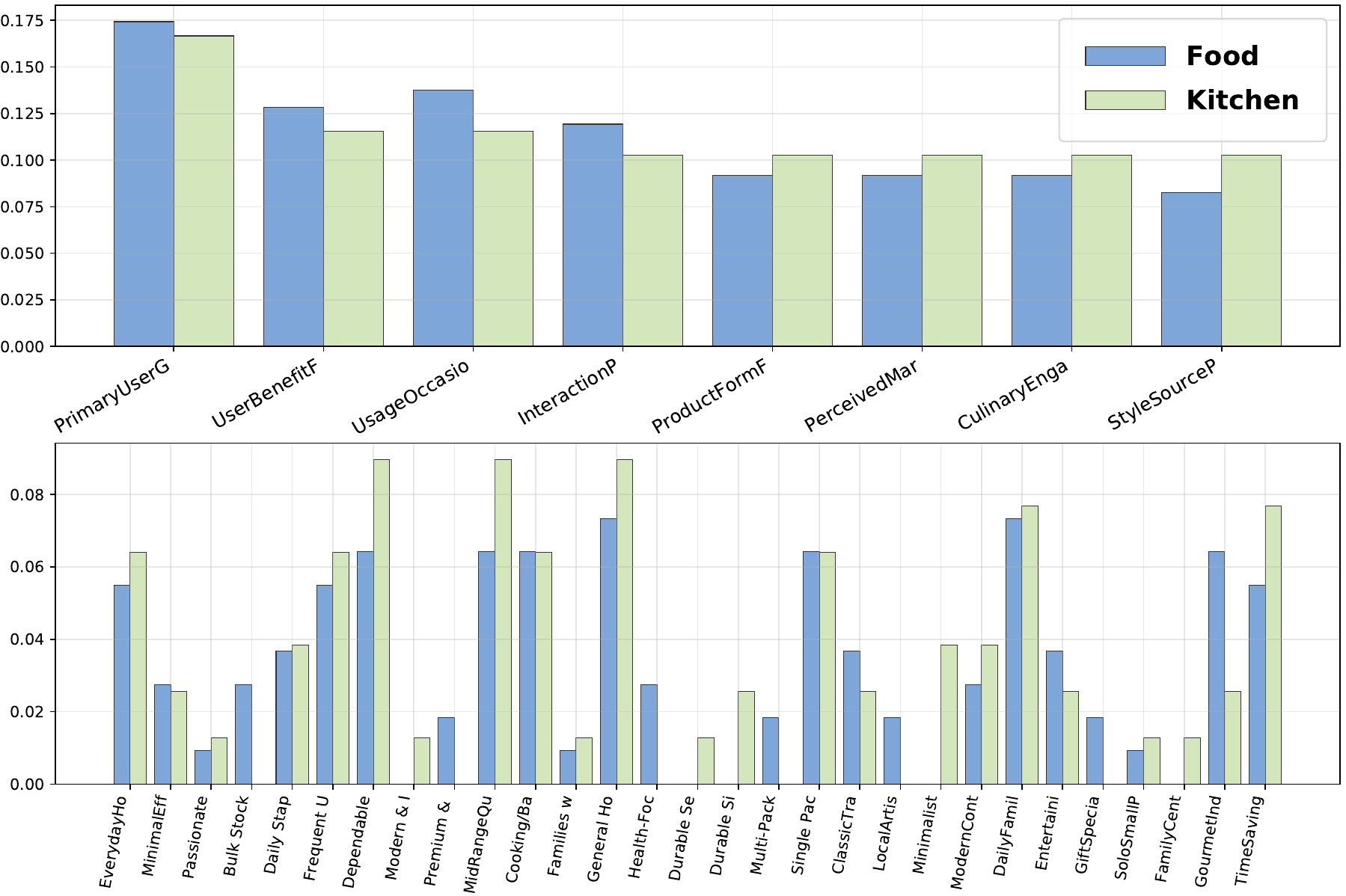}}}
\caption{User-level distributions of LLM-generated domain-agnostic semantic categories across source and target domains for Food–Kitchen dataset.}
    \label{model}
\end{figure}

\section{Baseline Descriptions}\label{app:baseline}
\textbf{Sequential recommendation models baselines:}
\begin{itemize}
    \item \textbf{SASRec}~\cite{kang2018self}. A sequence recommendation model uses the self-attention mechanism to capture temporal dependencies in users' behavior sequences, thus identifying and tracking their dynamically changing preferences.
    \item \textbf{CL4SRec}~\cite{xie2022contrastive}. A multi-task sequential recommendation model that jointly performs next-item prediction and contrastive learning, leveraging self-supervised signals to learn richer and more discriminative user representations.
\end{itemize}

\textbf{Cross-domain sequential recommendation baselines:}
\begin{itemize}
    \item \textbf{$\pi$-Net}~\cite{ma2019pi}. A pioneering cross-domain sequential recommendation model employing a Parallel Information-sharing Network, composed of a Shared Account Filter Unit (SFU) and a Cross-Domain Transfer Unit (CTU), to handle shared accounts and knowledge transfer across domains.
    \item \textbf{MGCL}~\cite{xu2025multi}. A model integrating intra- and inter-sequence item dependencies while simultaneously learning single- and cross-domain user preferences through a contrastive learning paradigm.
    \item \textbf{C$^2$DSR}~\cite{cao2022contrastive}. A model integrating intra- and inter-sequence item dependencies while simultaneously learning single- and cross-domain user preferences through a contrastive learning paradigm.
    \item \textbf{CD-SASRec}~\cite{alharbi2021cross}. An extension of SASRec for cross-domain scenarios, which transfers knowledge by incorporating aggregated source-domain representations into target-domain item embeddings.
    \item \textbf{Tri-CDR}~\cite{ma2024triple}. A framework utilizing source, target, and mixed behavior sequences with triple cross-domain attention and contrastive learning to model both comprehensive and domain-specific user preferences.
    \item \textbf{TJAPL}~\cite{xu2024transfer}. A cross-domain sequential recommendation model leveraging attentive preference learning to capture inter-preference correlations and transfer knowledge from multiple source domains to the target domain.
\end{itemize}

\section{Details about Hyper-parameter Analysis}\label{app:hyper}
\begin{figure}[htbp]
    \centering
    \setlength{\abovecaptionskip}{0.2mm}
    \includegraphics[width=\columnwidth]{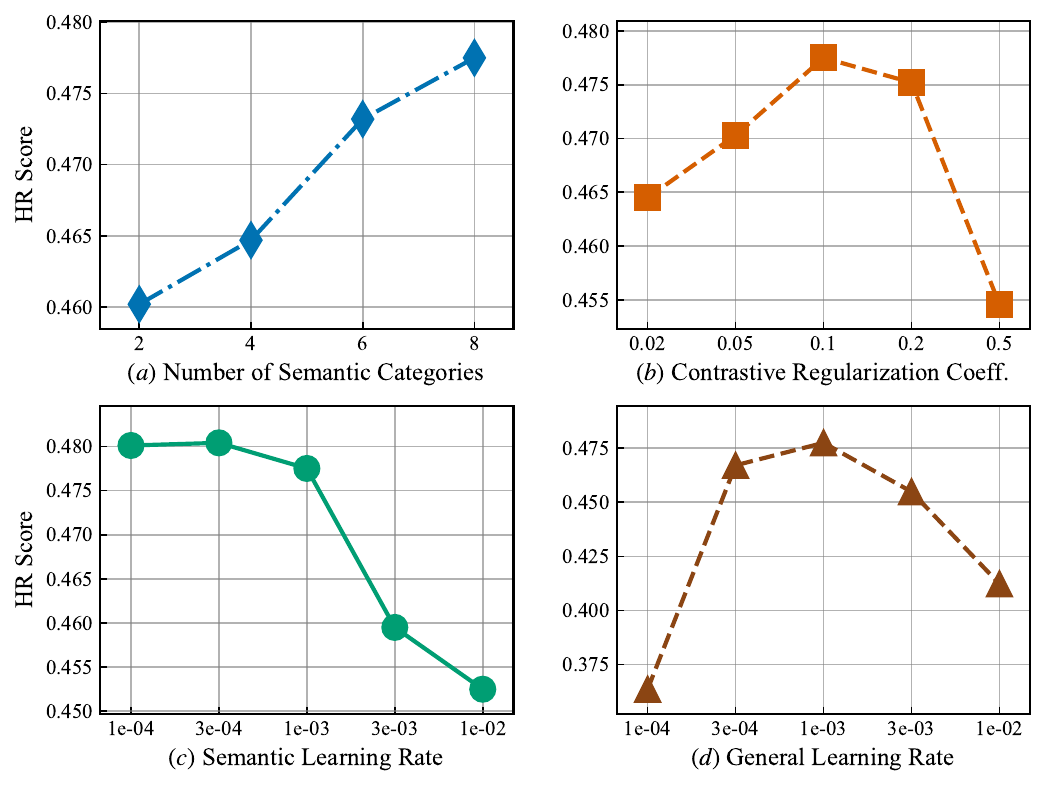} 
    \caption{Hyper-parameter analysis results.} 
    \label{fig:hyper_param}
\end{figure}

In this subsection, we further investigate the impact of several key hyperparameters on model performance and report the performance on the Kitchen-Food dataset in terms of HR@10. We first vary the number of domain-independent high-level semantic categories $K \in \{2, 4, 6, 8\}$. As shown in Figure~\ref{fig:hyper_param} (a), performance improves steadily as $K$ increases, since too few categories cannot adequately capture item characteristics. The best result is achieved at $K=8$, beyond which additional categories may introduce redundancy and unnecessary parameter overhead. This suggests that setting 
$K=8$ provides a balanced and effective representation.

We next analyze the coefficient $\lambda$ that weights the contrastive regularization between item views, balancing the primary recommendation objective with the contrastive signal. The results in Figure~\ref{fig:hyper_param} (b) indicate a clear trend: as $\lambda$ increases from 0.02 to 0.1, the performance steadily improves, peaking at $\lambda=0.1$ with an HR@10 of 0.4775. A smaller value provides insufficient contrastive guidance, whereas an excessively large value (\eg $\lambda=0.5$) disrupts optimization of the main recommendation task and leads to degradation. Based on these observations, we fix $\lambda=0.1$ in all subsequent experiments as a balanced and effective setting.

Finally, we analyze the effect of learning rate configuration on model performance. According to parameter functionality, we divide the model into two parts: semantic-related modules (for inner-domain semantic and domain-agnostic semantic embeddings) and general components, and assign them separate learning rates. As shown in Figure~\ref{fig:hyper_param} (c), (d), the best results are achieved when the semantic learning rate is set smaller than the general learning rate, with the semantic learning rate set to $3e^{-4}$ and the general learning rate set to $1e^{-3}$. This is consistent with intuition, since semantic representations are relatively stable and benefit from more conservative updates, while other components require larger adjustments to capture dynamic recommendation signals.

\end{document}